\newcommand{\myclearpage}{\clearpage}
\renewcommand{\myclearpage}{}

\newcommand{\ecoopOrExtended}[2]{#1}   
\renewcommand{\ecoopOrExtended}[2]{#2} 

\documentclass[cleveref,autoref,thm-restate]{lipics-v2021}

\ecoopOrExtended{}{\hideLIPIcs}
\nolinenumbers

\title{Mover Logic: A Concurrent Program Logic for Reduction and Rely-Guarantee Reasoning\ecoopOrExtended{}{ (Extended Version)\footnote{This is an extended version of a paper published at ECOOP 2024~\cite{DBLP:conf/ecoop/FlanaganF24}.}}} 

\titlerunning{Mover Logic} 

\author{Cormac Flanagan}{University of California, Santa Cruz, Santa Cruz, CA, USA}{}{0009-0009-5067-6774}{}
\author{Stephen N. Freund}{Williams College, Williamstown, MA, USA}{freund@cs.williams.edu}{0009-0000-6992-199X}{}

\authorrunning{C. Flanagan and S.N. Freund} 

\Copyright{Cormac Flanagan and Stephen N. Freund} 

\ccsdesc[500]{Theory of computation~Program verification}
\ccsdesc[500]{Theory of computation~Program specifications}
\ccsdesc[500]{Software and its engineering~Concurrent programming languages}
\ccsdesc[500]{Software and its engineering~Formal software verification}

\keywords{concurrent program verification, reduction, rely-guarantee reasoning, synchro\-nization} 

\EventEditors{Jonathan Aldrich and Guido Salvaneschi}
\EventNoEds{2}
\EventLongTitle{38th European Conference on Object-Oriented Programming (ECOOP 2024)}
\EventShortTitle{ECOOP 2024}
\EventAcronym{ECOOP}
\EventYear{2024}
\EventDate{September 16--20, 2024}
\EventLocation{Vienna, Austria}
\EventLogo{}
\SeriesVolume{313}
\ArticleNo{3}

\funding{This work was supported by NSF grants 2243636 and 2243637.}

\usepackage{booktabs}   
\usepackage{subcaption} 
\usepackage[export]{adjustbox}
\usepackage{listings}
\newcommand{\threadonecol}[0]{red}

\definecolor{\threadonecol}{RGB}{177,0,28}

\newcommand{\inlinecode}[1]{\lstinline[basicstyle=\tt\normalsize]{#1}}

\newcommand{\logic}[0]{mover logic}
\newcommand{\Logic}[0]{Mover logic}
\newcommand{\LOGIC}[0]{Mover Logic}

\begin{document}

\maketitle

\begin{abstract}

Rely-guarantee (RG) logic uses thread interference specifications
(relies and guarantees) to reason about the correctness of
multithreaded software.  Unfortunately, RG logic requires each
function postcondition to be ``stabilized'' or specialized to the
behavior of other threads, making it difficult to write function
specifications that are reusable at multiple call sites.

This paper presents mover logic, which extends RG logic to address
this problem via the notion of atomic functions.  Atomic functions
behave as if they execute serially without interference from
concurrent threads, and so they can be assigned more general and
reusable specifications that avoid the stabilization requirement of
RG logic.  
Several practical verifiers (Calvin-R, QED, CIVL, Armada, Anchor, etc.)
have demonstrated the modularity benefits of atomic function
specifications.  However, the complexity of these systems and their
correctness proofs makes it challenging to understand and extend these systems.
Mover logic formalizes the central ideas reduction in a declarative program logic that may provide foundation for future work in this area.
\end{abstract}

\maketitle

\newcommand{\qsep}{.\;}
\renewcommand{\note}[1]{\mbox{\textcolor{red}{[#1]}}}

\newcommand{\relsize}{\small}
\newcommand{\relname}[1]{\raisebox{0ex}[0ex][0ex]{\mbox{\relsize \textsc{[#1]}}}}
\newcommand{\myheadrule}[1]{\fbox{\ensuremath{#1}}\vspace{1mm}}

\newcommand{\rln}[2]{
\mbox{
$\begin{array}{@{}l}
\frac 
    {\strut\begin{array}{c} #1 \end{array}} 
    {\strut\begin{array}{c} #2 \end{array}}
\end{array}
$}}

\newcommand{\nrln}[3]{
\mbox{
$\begin{array}{@{}l@{}}
\relname{#1} \\
~~\frac 
    {\strut\begin{array}{c} #2 \end{array}} 
    {\strut\begin{array}{c} #3 \end{array}}
\end{array}
$}}

\newcommand{\nrlnb}[2]{
\mbox{
$\begin{array}{@{}l@{}}
\relname{#1} \\
    {\strut\begin{array}{c} #2 \end{array}}
\end{array}
$}}

\newcommand{\nrlnp}[3]{
\relname{#1} &
  \frac 
    {\strut\begin{array}{c} #2 \end{array}} 
    {\strut\begin{array}{c} #3 \end{array}}
}

\newcommand{\nrlnpp}[3]{
\relname{#1} \quad
  \frac 
    {\strut\begin{array}{c} #2 \end{array}} 
    {\strut\begin{array}{c} #3 \end{array}}
}

\newcommand{\nrlnnoname}[2]{
  \frac 
    {\strut\begin{array}{c} #1 \end{array}} 
    {\strut\begin{array}{c} #2 \end{array}}
}

\newcommand{\tword}[1]{
  \expandafter\def\csname #1\endcsname{\mbox{{\tt #1}}}}
\newcommand{\ttword}[2]{
  \expandafter\def\csname #1\endcsname{\mbox{{\tt #2}}}}

\newcommand{\bfword}[1]{
  \expandafter\def\csname b#1\endcsname{\mbox{{\bf #1}}}}

\newcommand{\fs}[1]{\mathit{#1}}

\renewcommand{\t}[1]{\texttt{#1}}

\newcommand{\mover}[0]{e}

\ttword{tskip}{skip}
\ttword{tyield}{yield}
\ttword{twrong}{wrong}
\ttword{tif}{if}
\ttword{twhile}{while}
\ttword{tatomic}{atomic}
\ttword{tinatomic}{inatomic}
\ttword{telse}{else}

\bfword{atomic}
\bfword{requires}
\bfword{guarantees}
\bfword{relies}
\bfword{ensures}
\bfword{moves}

\newcommand{\defeq}{\stackrel{{\mbox{def}}}{=}}

\newcommand{\Tid}[0]{\mathit{Tid}}
\newcommand{\at}[0]{t}
\newcommand{\au}[0]{u}
\newcommand{\tid}[0]{\fs{tid}}
\newcommand{\Store}[0]{\mathit{Store}}
\newcommand{\State}[0]{\mathit{State}}
\newcommand{\IState}[0]{\mathit{IState}}
\newcommand{\Var}[0]{\mathit{Var}}
\newcommand{\Value}[0]{\mathit{Value}}

\newcommand{\sif}[3]{\tif ~ #1 ~ #2 ~ \telse ~ #3}
\newcommand{\swhile}[2]{\twhile ~ #1 ~ #2}
\newcommand{\seq}[2]{#1; #2}
\newcommand{\siter}[1]{#1^*}
\newcommand{\satomic}[1]{\tatomic~#1}
\newcommand{\sinatomic}[3]{\tinatomic^{#1} ~ #2; #3}
\newcommand{\action}[2]{#1^{#2}}
\newcommand{\dpar}[2]{#1\, |\!|\, #2}

\newcommand{\batom}[1]{\textcolor{blue}{\texttt{\textbf{#1}}}}
\newcommand{\MN}[0]{{\batom{N}}}
\newcommand{\MB}[0]{{\batom{B}}}
\newcommand{\ML}[0]{{\batom{L}}}
\newcommand{\MY}[0]{{\batom{Y}}}
\newcommand{\MR}[0]{{\batom{R}}}
\newcommand{\ME}[0]{{\batom{E}}}

\newcommand{\preemptive}[2]{#1 \rightarrowtail #2 }
\newcommand{\nonpreemptive}[2]{#1 \mapsto #2 }

\newcommand{\pstate}[2]{#1 \cdot #2}
\newcommand{\istate}[3]{#3\! \cdot\! #1\! \cdot\! #2}

\newcommand{\steps}[2]{#1 \leadsto #2 }
\newcommand{\stepsp}[2]{#1 & \leadsto & #2 }
\newcommand{\threadstep}[5]{#1 \vdash \pstate{#2}{#3}  \leadsto \pstate{#4}{#5}}
\newcommand{\rgmimp}[6]{#1;#2;#3 \vdash #4 : #5 \rightarrow #6}
\newcommand{\rtimp}[6]{#1;#2;#3 \vdash #4 : #5 \cdot #6}
\newcommand{\impstate}[1]{\vdash #1}

\newcommand{\post}[1]{\fs{Post}(#1)}
\newcommand{\two}[1]{\fs{Two}(#1)}
\newcommand{\yield}[2]{\fs{Yield}(#1, #2)}

\newcommand{\mimp}[6]{#1;#2 \vdash #3 : #4 \rightarrow #5 \cdot #6}
\newcommand{\fspec}[5]{
        {\bf relies}~#1 
        {\bf guarantees}~#2 
        {\bf moves}~#3 
        {\bf requires}~#4
        {\bf ensures}~#5
        }
\newcommand{\fspeca}[5]{
        {\bf relies}~#1 \\
        {\bf guarantees}~#2 \\
        {\bf moves}~#3  \\
        {\bf requires}~#4 \\
        {\bf ensures}~#5
  }

\newcommand{\faspec}[4]{
        {\bf atomic}~#1~
        {\bf moves}~#2~
        {\bf requires}~#3~
        {\bf ensures}~#4
        }
\newcommand{\faspeca}[4]{
        {\bf atomic}~#1 \\
        {\bf moves}~#2 \\
        {\bf requires}~#3\\
        {\bf ensures}~#4 
        
  }

\newcommand{\boolcond}[0]{C}
\newcommand{\boolaction}[0]{C}
\newcommand{\teq}{\t{\textasciitilde=}}

\section{Introduction}


Verifying that a multithreaded software system behaves correctly for
all possible inputs and thread interleavings is a critically important
problem in computer science.  To verify large systems, verification
techniques must employ \emph{modular reasoning} in which each
function's implementation is verified with respect to its
specification.
In a multithreaded system, writing precise and reusable function
specifications is a rather difficult challenge, since concurrent
threads can observe and change the state of a function call not just
in its initial and final states, but also at any intermediate states
during the function's execution.  Thus, function specifications must
describe not just the function's precondition and postcondition, but
also how the function may influence and be influenced by other
concurrent threads.
To address this  problem,
Rely-Guarantee (RG) logic~\cite{DBLP:journals/toplas/Jones83} uses
function
specifications  that include:

\begin{itemize}
  \item a \emph{guarantee} $G$  describing how each step of
the function may update shared state, and 
  \item a \emph{rely assumption} $R$  describing the behavior of interleaved steps of other
threads. The rely assumption might, for example, specify that
interleaved steps preserve a data invariant.
\end{itemize}

%
Under RG logic, however, a function's postcondition must summarize not
only the behavior the function itself but also the behavior of
interleaved steps of other
threads~\cite{DBLP:conf/esop/WickersonDP10}.  Consequently, RG
function specifications are often specialized to the rely assumption
and data invariants of a particular client, limiting reuse of those
function specifications in other clients, as we illustrate in
Section~\ref{sec:overview}.

Lipton's theory of reduction~\cite{DBLP:journals/cacm/Lipton75}
provides a promising approach to address this problem.  It uses a
commuting argument to show that certain functions are \emph{atomic}
and behave {as if} they execute serially (without interleaved
steps of other threads). Consequently, atomic functions do not require
interleaved rely assumptions, and they can be precisely specified
using preconditions and postconditions that are independent of any
specific client.

Reduction has been widely adopted in a variety of software validation
tools, including dynamic
analyses~\cite{DBLP:conf/popl/FlanaganF04,DBLP:conf/ppopp/WangS06,DBLP:journals/tse/WangS06,DBLP:conf/fase/ChenWYS09},
type
systems~\cite{DBLP:conf/ppopp/SasturkarAWS05,DBLP:conf/tldi/FlanaganQ03,DBLP:conf/pldi/FlanaganQ03,DBLP:conf/issta/FlanaganFQ04},
and other
tools~\cite{DBLP:journals/fmsd/CernyCHRRST17,DBLP:journals/scp/YiDFF15,DBLP:conf/tldi/YiF10,DBLP:conf/issta/YiDFF12}. Over
the past two decades, software verifiers based on reduction (\emph{e.g.}, Calvin-R~\cite{DBLP:journals/jot/FreundQ04},
QED~\cite{DBLP:conf/icse/Elmas10},
CIVL~\cite{DBLP:conf/cav/HawblitzelPQT15,DBLP:conf/fmcad/KraglQ21},
Armada~\cite{DBLP:conf/pldi/LorchCKPQSWZ20}, and
Anchor~\cite{DBLP:journals/pacmpl/FlanaganF20}) have
demonstrated the utility of atomic function specifications in
verifying sophisticated concurrent code.
%
%
%
To date, however, reduction-based verifiers have not been based on an
underlying program logic, such as RG logic. Instead, their soundness
arguments are typically based on monolithic proofs whose complexity
inhibits further research.
To address this complexity barrier, we present mover logic,
which extends RG logic to support atomic function specifications via
reduction-based reasoning.

In mover logic,
%
thread interference points are documented with \tyield{}
annotations that have no run-time effect.
Mover logic verifies that every sequence of operations between two
yield points is reducible and hence amenable to sequential reasoning.
In order to verify reducibility, mover logic uses synchronization
specifications describing both when each thread can access each shared
location and how those accesses commute with concurrent accesses of
other threads. 
In contrast to RG logics that must stabilize all state predicates
under the rely assumption, mover logic only needs to stabilize
predicates at \tyield{} points.
%
%
Atomic functions have no yield points and can thus be specified with
traditional pre- and postconditions.  Moreover, 
atomic function specifications need not include a client-specific rely
assumption that would limit reuse in other clients that have different
rely assumptions or data invariants.

%
Mover logic is a declarative program logic (similar in style to Hoare
Logic and RG Logic) that helps explain
and justify many subtle aspects of reduction-based verification,
including:
\begin{itemize}
  \item what code blocks are reducible;
  \item where yield annotations are required;
  \item which functions are atomic; 
  \item what atomic and non-atomic function specifications mean;
  \item what reasoning is performed by the verifier; 
  \item why this reasoning is sound; and
  \item which programs are verifiable or not verifiable, and why.
\end{itemize}
Mover logic simplifies the soundness proof for any
specific verifier, because the proof now must only show that the verifier follows
the rules of mover logic.

\medskip
\noindent
The presentation of our results proceeds as follows.
\begin{itemize}
\item Section~\ref{sec:overview} illustrates the specification entanglement
  problem of RG logic and shows how mover logic avoids this problem.
  
\item Section~\ref{sec:reduction} reviews Lipton's theory of reduction. 
  
\item Sections~\ref{sec:ml-overview} and~\ref{sec:examples} present an overview of mover logic and additional   examples. 
  
\item Section~\ref{sec:lang}  formalizes a core multithreaded language.
  
\item Section~\ref{sec:effects} and~\ref{sec:logic} present mover logic for this language. 
  
\item Sections~\ref{sec:related} and~\ref{sec:summary} discuss
  related work and summarize our contributions.
\end{itemize}

For clarity of exposition, our presentation of mover logic targets an idealized multithreaded language that captures the essential
complexities of multithreaded function specifications.  Extending the
logic to more complex languages remains an important topic for future
work.

\myclearpage

\section{Limitations of Rely-Guarantee Logic}
\label{sec:overview}


\begin{figure}[p!]
  \includegraphics[width=0.90\textwidth]{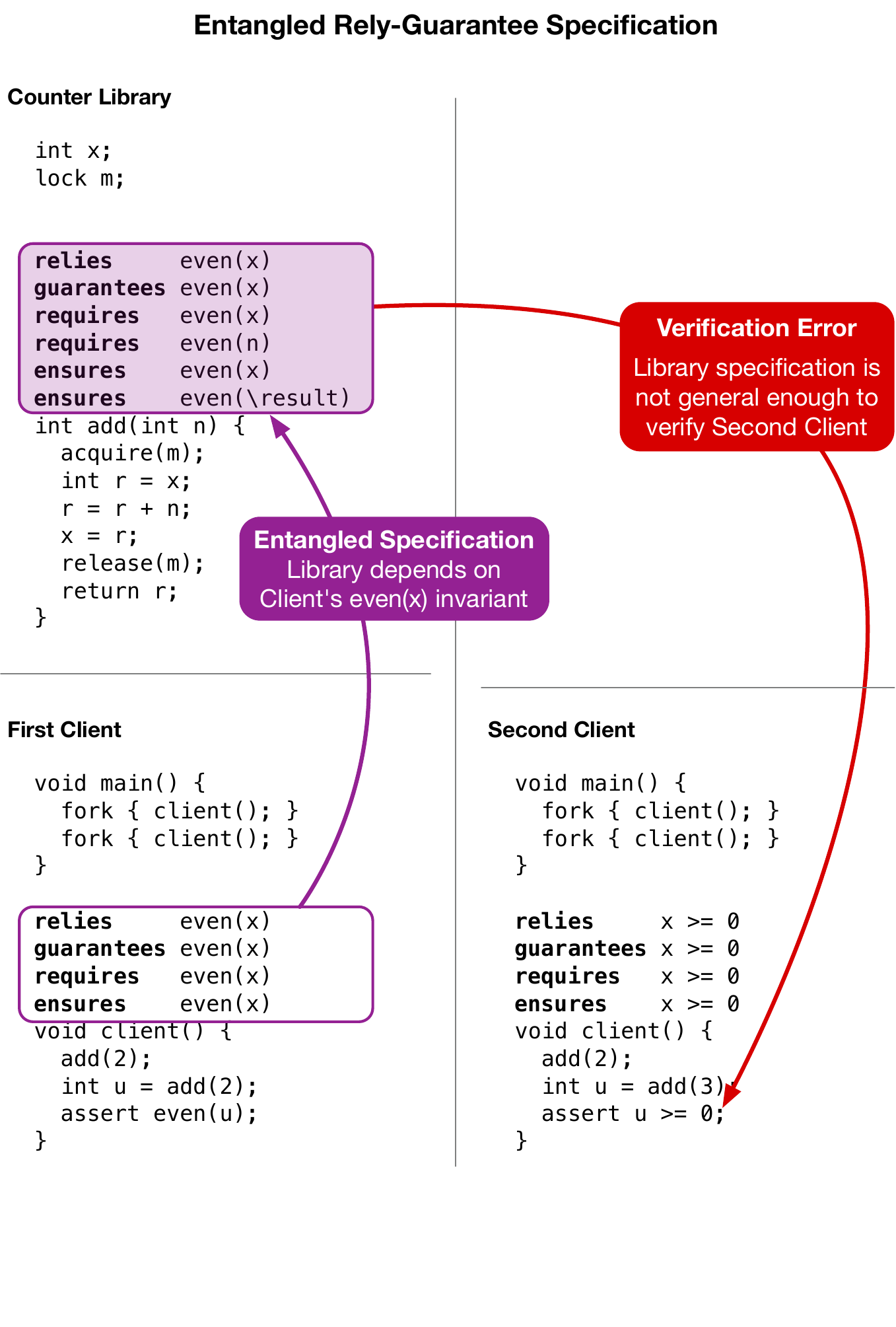}
  \vspace*{-0.75in}
  \caption{\label{fig:set}
    Our idealized running example is an
    \t{add(n)} library function that atomically increases shared
    variable \t{x} by \t{n}.
    %
    %
    \textbf{(Left)} A rely-guarantee specification.  The client's data
    invariant \t{even(x)} becomes entangled in the library
    specification. \textbf{(Right)} A second client that cannot be
    verified because the specification is
    insufficiently general.}
  \end{figure}

We motivate the need for mover logic via the example code in Figure~\ref{fig:set}~(left).  That code consists of:
\begin{enumerate} 
\item A \textbf{counter library} that contains the function \t{add(n)} that adds
  \t{n} to the variable \t{x} and returns the new value
  of \t{x}.  The initial value of variable \t{x} is 0, and it is protected by lock
  \t{m}, whose value is either the thread identifier $tid$ of the thread holding the lock or 0 if unheld.  The lock is initially unheld.   
\item A \textbf{first client} that creates two threads, and each thread calls \t{add(2)}
  multiple times before asserting that \t{x} is even.
\end{enumerate}
This program verifies under RG logic based on the
invariant that \t{x} is always even.
This \t{even(x)} invariant is a precondition and postcondition for
both \t{add()} and \t{client()}\footnote{Frame conditions, which
specify the locations a function may read or modify, also play a key
role in modular function specifications, but we do not consider them in this paper due to lack of space.  Extending mover logic with frame conditions, perhaps using ideas from separation logic~\cite{DBLP:conf/csl/OHearnRY01,DBLP:conf/lics/Reynolds02}, remains an important topic for future work.}:
\begin{center}
\begin{tabular}{c}
\begin{lstlisting}[basicstyle=\tt]
requires even(x)
ensures  even(x)
\end{lstlisting}
\end{tabular}
\end{center}
In addition, each step by each thread in the program is guaranteed to
preserve this invariant.  As a result, each thread can rely on other
threads to preserve the invariant:
\begin{center}
\begin{tabular}{c}
\begin{lstlisting}[basicstyle=\tt]
relies     even(x)
guarantees even(x)	
\end{lstlisting}
\end{tabular}
\end{center}
%


\begin{figure}[p!]
\includegraphics[width=\textwidth]{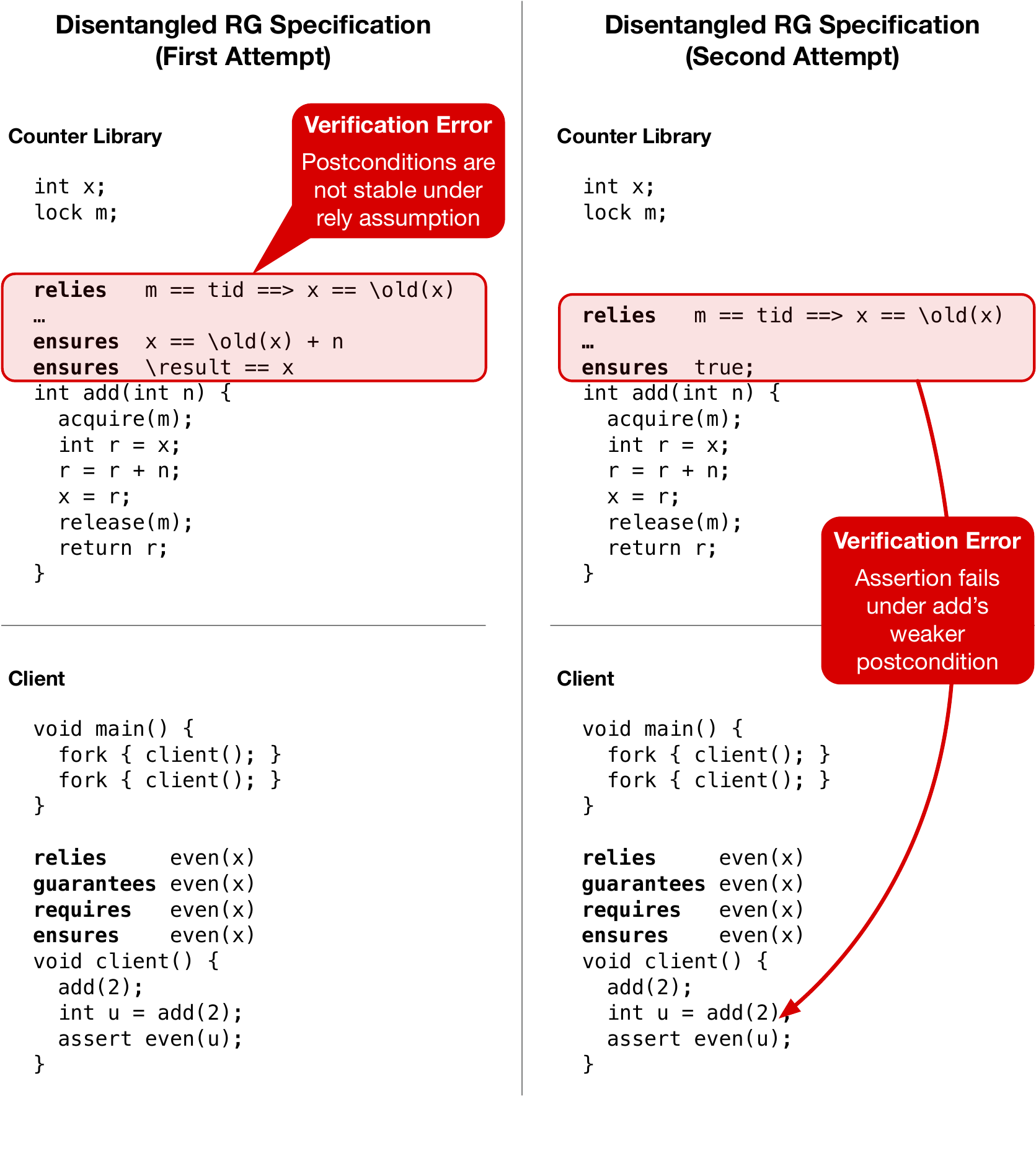}
\caption{\label{fig:set-failed} \textbf{(Left)} An attempt to
  disentangle the library specification from the client that does not
  meet RG stability requirements.  \textbf{(Right)} Another attempt
  that meets stability requirements but fails to verify the client.}
\end{figure}


These RG specifications are sufficient to verify that the program does
not go wrong by failing the \t{even(u)} assertion in \t{client()}, but
unfortunately the specification for \t{add()} is tightly-coupled, or
\emph{entangled}, with the \t{even(x)} data invariant from this
particular client.  A different client would necessitate revising and
re-verifying the counter library, which makes 
modular verification more challenging and less scalable.  For example,
the second client in Figure~\ref{fig:set}~(right) enforces the
data invariant $\t{x >= } 0$, but it cannot be verified with the
\t{add()} specification entangled with the first client.  Others have
noted this limitation as well (see, for
example,~\cite{DBLP:conf/esop/DoddsFPV09,DBLP:conf/esop/WickersonDP10}).

\subsection{Disentangling RG Specifications: First Attempt}

The goal of this paper is to support specifications for library functions like
\t{add()} that are \emph{not} specialized to one particular client.
As a first attempt to achieve that goal, the code in
Figure~\ref{fig:set-failed}~(left) uses the following natural
postconditions for \t{add()}, where \t{\textbackslash old(x)} and \t{x} refer to
the value of \t{x} upon function entry and exit, respectively:
\begin{center}
\begin{tabular}{c}
\begin{lstlisting}[basicstyle=\tt]
ensures x == \old(x) + n
ensures \result == x
\end{lstlisting}
\end{tabular}
\end{center}
In addition, if \t{add()} has no knowledge of its client, it must
assume that other client threads could call \t{add()} with arbitrary
arguments at any time, and so the natural rely assumption is that
other threads may update \t{x} whenever the lock \t{m} is not held by
the current thread.  That assumption is most easily expressed as its
contra-positive (where \t{tid} is the identifier of the current thread
and lock \t{m} is held by that thread when \t{m == tid}):
\begin{center}
\begin{tabular}{c}
\begin{lstlisting}[basicstyle=\tt]
relies  m == tid ==> x == \old(x)
\end{lstlisting}
\end{tabular}
\end{center}
Here,
\t{\textbackslash old(x)} and \t{x} refer to the value
of \t{x} before and after an interleaved action of another thread,
respectively.

To account for interleaved steps of other threads, a central
requirement of RG logic is that all store predicates (\emph{e.g.}
preconditions, postconditions, and invariants) used to reason about
program behavior must be \emph{stable} under this rely assumption $R$.
This means that interleaved $R$-steps from other threads must not
invalidate those predicates.
%
%
In the case of \t{add} in Figure~\ref{fig:set-failed}~(left), the postcondition \t{x == \textbackslash old(x) + n \&\& \textbackslash result == x} is not stable under the rely
assumption $R$, reflecting that \t{x} could be concurrently modified after
the lock is released but before \t{add()} returns.  Thus,
Figure~\ref{fig:set-failed}~(left) does not verify under RG logic.

\subsection{Disentangling RG Specifications: Second Attempt}

To ensure stability we must weaken the \t{add()} function's postcondition to be
stable under the rely assumption, as shown in
Figure~\ref{fig:set-failed}~(right).  Unfortunately, the resulting stable post
condition is simply \t{true}, which no longer guarantees anything about the value of
\t{x} and is too weak to verify the client.

\begin{figure}[tp!]
  \begin{center}
	\includegraphics[width=\textwidth]{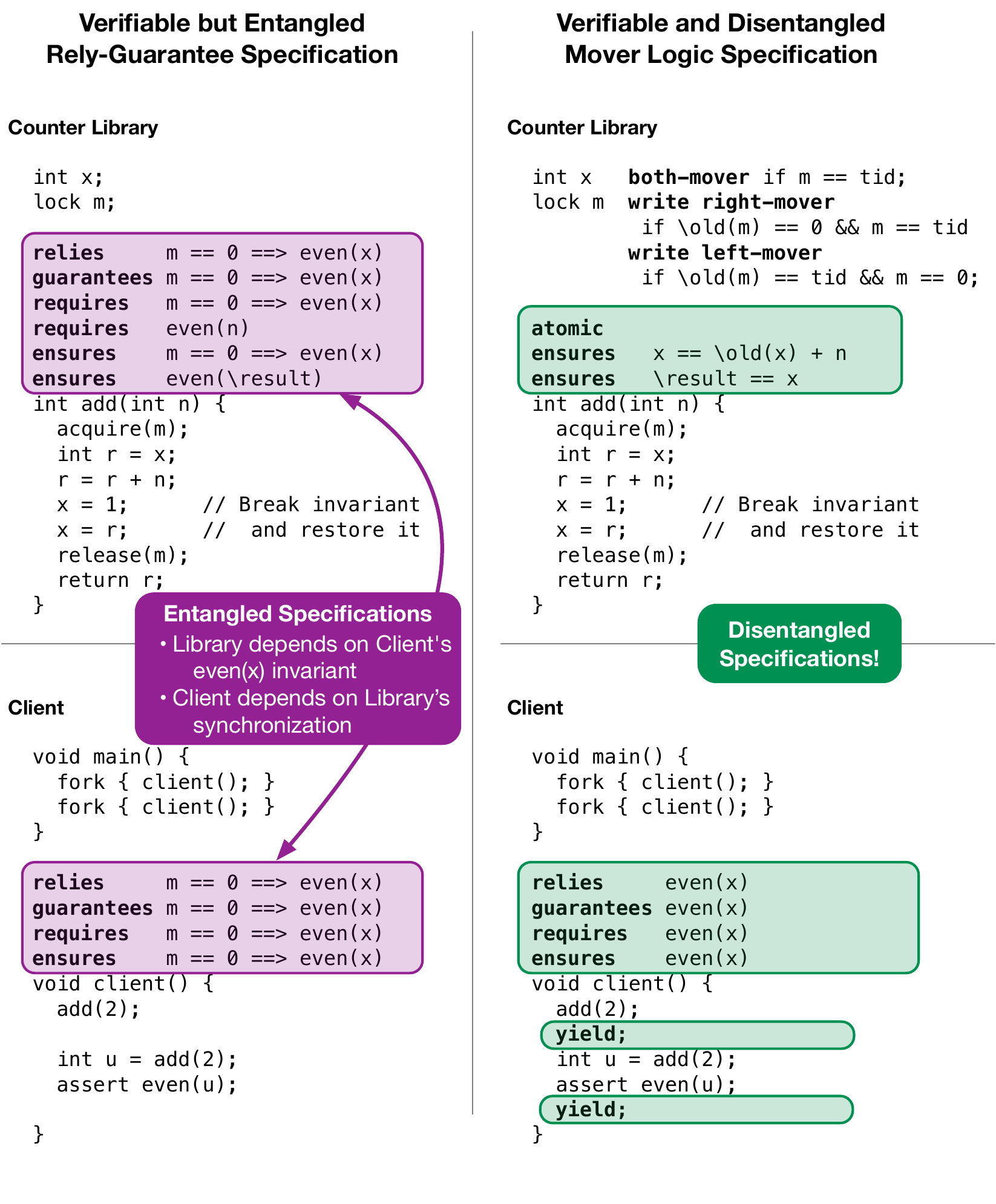}
  \end{center}
\caption{\label{fig:set-inv} A second version of the counter library
  with a temporarily broken $\t{even(x)}$ invariant. \textbf{(Left)}
  Under RG logic, the library specification is entangled with the
  client's \t{even(x)} invariant and the client specification is
  entangled with the library's synchronization discipline.
  \textbf{(Right)} Under \logic, the specifications are cleanly disentangled.}
\end{figure}

\subsection{Broken Invariants and Bidirectional Entanglement}

As a more challenging example, consider the \t{add()} library variant
in Figure~\ref{fig:set-inv}~(left) that temporarily breaks the
\t{even(x)} invariant while holding the lock.  In this case, the
invariant holds only when lock \t{m} is free:
\begin{center}
\begin{tabular}{c}
\begin{lstlisting}[basicstyle=\tt]
m == 0 ==> even(x)
\end{lstlisting} 
\end{tabular}
\end{center}
Stores at the program points in which the invariant is broken are not
intended to be observable by clients.  However, the revised RG
specifications for \t{add()} and the client must now be based on
this conditional invariant, resulting in two problems.  First, the
library specification is again specialized to the client's \t{even(x)}
invariant.  Second, the library's internal locking discipline leaks
into the client's specification,
limiting our ability to modify the library code without
breaking clients.
This example demonstrates that RG reasoning may force us to lose
modularity between client and library.


\begin{figure}[t!]
  \begin{center}
  \includegraphics[width=1\textwidth]{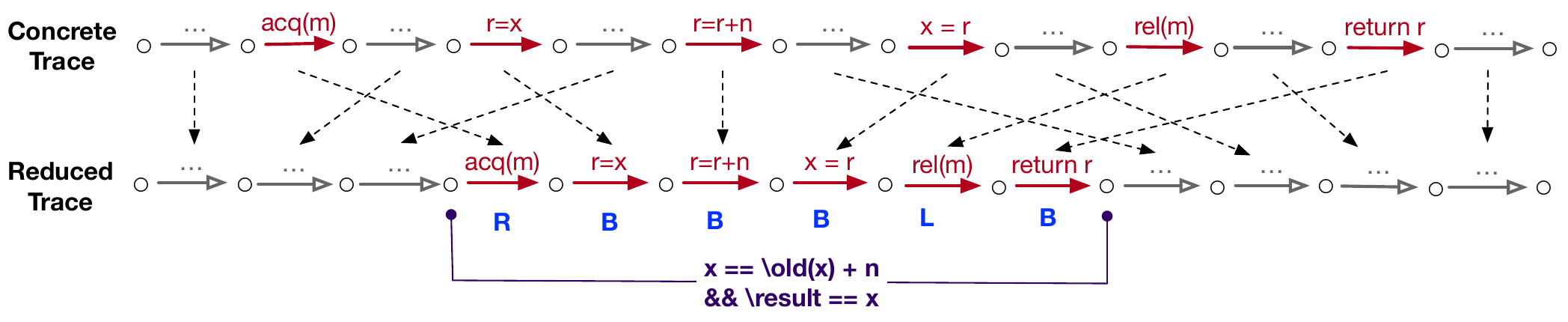}
  \end{center}
  \caption{\label{fig:red} Reduction applied to an execution trace of \t{add()} from Figure~\ref{fig:set}.}
\end{figure}

\section{Review of Lipton's Theory of Reduction\label{sec:reduction}}

Our solution to this specification problem employs Lipton's
theory of reduction~\cite{DBLP:journals/cacm/Lipton75}, which 
classifies how steps of one thread commute with concurrent
steps of another thread.
\begin{itemize}
  \item A step is a \emph{right-mover} ($\MR$) if it commutes ``to the
    right'' of any subsequent step by a different thread, in that
    performing the steps in the opposite order does not change the
    final store.  A lock acquire is a right-mover because any
    subsequent step from another thread cannot modify that lock.

\item Conversely, a step is a \emph{left-mover} ($\ML$) if it commutes
  ``to the left'' of a preceding step of a different thread.  A lock
  release is a left-mover because any preceding step cannot modify
  that lock.

\item A step is a \emph{both-mover} ($\MB$) if it is both a
  left- and a right-mover, and it is a \emph{non-mover} ($\MN$) if
  neither.  A race-free memory
  access is a both-mover because there are no concurrent,
  conflicting accesses.  An access to a race-prone variable is a
  non-mover since there may be concurrent writes.
\end{itemize}
A sequence of steps performed by a particular thread is
\emph{reducible} if consists of
  (1) zero or more right-movers;
  (2) at most one non-mover; and
  (3) zero or more left-movers.  That is, a sequence is reducible if
the commutativity of the steps match the pattern $\MR^* [\MN] \ML^*$.
Any interleaved steps of other threads can be ``commuted out'' to
produce a serial execution. 

Figure~\ref{fig:red}
illustrates this technique for a call to \t{add()} interleaved with
steps of a second thread.  In that figure and below, the solid and
hollow arrow heads indicate steps from different threads, and arrows
labeled ``\ldots'' represent any number of steps by that thread.  The
steps of \t{add()} have the mover behavior $\MR\, \MB\, \MB\, \MB\,
\ML\, \MB$, matching the reducible pattern $\MR^* [\MN] \ML^*$.  Thus
we can reason about \t{add()} as if it executes sequentially and
assign it the intuitive postcondition $\verb{x == \old(x) + n && \result == x{$.

\section{Overview of \LOGIC}
\label{sec:ml-overview}

{\Logic} extends RG logic to verify that certain functions are
\emph{atomic} and can therefore be assigned more precise
(unstabilized) postconditions than under RG logic.
Figure~\ref{fig:set-inv}~(right) shows a mover logic specification for
our library/client example. The declaration
\begin{center}
\begin{tabular}{c}
\begin{lstlisting}[basicstyle=\tt]
int x  both-mover if m == tid;
\end{lstlisting}
\end{tabular}
\end{center}
means that accesses to \t{x} are both-movers provided that the current
thread holds lock \t{m}.  All other accesses are errors.  The
declaration for lock \t{m} specifies that acquires (which change \t{m}
from 0 to the current thread's identifier \t{tid}) are right-movers
and releases (which change \t{m} from \t{tid} back to 0) are
left-movers:
\begin{center}
\begin{tabular}{c}
\begin{lstlisting}[basicstyle=\tt]
lock m  write right-mover 
         if \old(m) == 0 && m == tid
        write left-mover 
         if \old(m) == tid && m == 0;
\end{lstlisting}
\end{tabular}
\end{center}
These mover specifications are sufficient to verify that
\t{add} is atomic.
Consequently, there is no need to apply the rely
assumption at each intermediate store inside this atomic
function. Instead, sequential reasoning suffices to establish
the desired postcondition $\verb{x == \old(x) + n && \result == x{$.

The \t{client()} function in Figure~\ref{fig:set-inv}~(right) is not
atomic because steps of other threads could be interleaved between the
two calls to \t{add(2)}.  Mover logic uses a $\tyield$ annotation to
identify that thread interference may occur at that point, and the
store invariants at {\tyield}s must be stable under the rely
assumption:
\begin{center}
\begin{tabular}{c}
\begin{lstlisting}[basicstyle=\tt]
relies      even(x);	
guarantees  even(x);
\end{lstlisting}
\end{tabular}
\end{center}
Note that this thread guarantee does not need to summarize individual
steps inside the callee \t{add()}, which would expose the broken
invariant.  Instead, it summarizes the entire atomic effect of
\t{add()}, which preserves the \t{even(x)} invariant.  With mover
logic, the \t{client()} specification is independent of the
internal synchronization discipline inside \t{add()}.
 
\smallskip
\noindent
This library/client example illustrates several benefits of
mover logic:
\begin{itemize}
  \item Verifying that \t{add()} is atomic enables
    sequential reasoning inside \t{add()}.
  \item We thus avoid applying the rely assumption at each
    program point inside \t{add()}.
  \item As a result, \t{add()} satisfies the desired postcondition
    $\verb{x == \old(x) + n && \result == x{$, which is
    independent of the client-specific data invariant
    \inlinecode{even(x)}.
  \item On the client side, the thread guarantee  \inlinecode{even(x)} summarizes the entire
    behavior of \t{add()}, rather than the behavior of each individual
    step.
  \item Consequently, the client can be verified based on the illusion
    that \t{even(x)} always holds, with no loss of soundness.
\end{itemize}
Thus, mover logic disentangles the library
specification from the data invariant of the client while also
disentangling the client specification from the library
synchronization discipline.

\begin{figure}[p!]
  \includegraphics[width=1\textwidth]{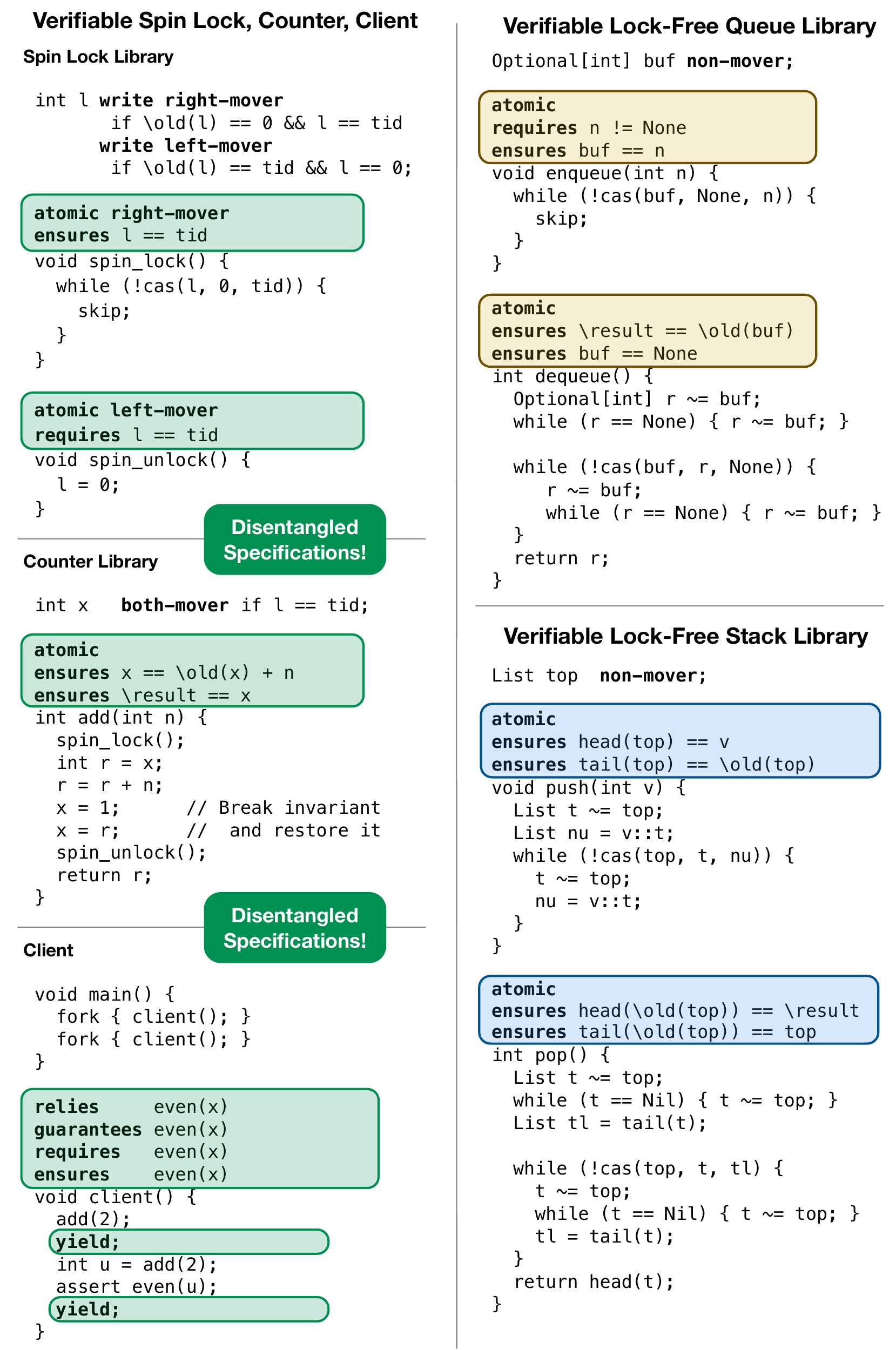}
  \caption{\label{fig:spin} \textbf{(Left)} A new implementation of
    the counter library using a user-defined spin
    lock. 
    \textbf{(Top Right)} A single-element lock-free queue.
    \textbf{(Bottom Right)} A lock-free stack. 
    }
\end{figure}


\newcommand{\sdots}[2]{#1..#2}

\newcommand{\eeval}[5]{
\relname{#1} & #3 & \rightarrow_{#2} & #4 & #5
}

\newcommand{\ieval}[4]{
\relname{#1} \!\!\!\!\!\!\!\!\!\!\!\!\!\! & #2 & \rightarrow_{t} & #3 & #4 
}

\newcommand{\iwhile}[3]{
\relname{#1} \!\!\!\!\!\!\!\!\!\!\!\!\!\!  & #2 & \rightarrow_{t} & \multicolumn{2}{l}{#3}
}

\newcommand{\dotact}[2]{#1\! \diamond \! #2}

\section{Additional Examples}
\label{sec:examples}

\subsection{Spin Lock}
\label{sec:spin}

To further illustrate the benefits of disentangled specifications,
Figure~\ref{fig:spin} (left) shows our counter library rewritten to employ
a user-defined spin lock.  The \t{spin\_lock()} code employs a
compare-and-set operation (\t{cas}) to attempt to change the lock
\t{l} from 0 to the current thread's \t{tid}.  The \t{cas} operation
returns true if the update succeeds, and false otherwise.  Thus, the
function retries until the update is success, at which point the
current thread holds the lock.  The \t{spin\_unlock()} function
releases the lock by setting \t{l} back to 0.

Mover logic verifies that calls to \t{spin\_lock()} and
\t{spin\_unlock()} are atomic right- and left-movers, respectively.
That enables us to avoid entangled specifications for the spin lock
and counter libraries, and the counter library's \t{add} specification
is \emph{identical} to the earlier implementation.  It is still atomic
and it guarantees the same post condition.

\subsection{Lock-Free Queue}
\label{sec:queue}

Figure~\ref{fig:spin} (top right) shows a lock-free single-element queue, where \t{buf} holds either the single enqueued \t{int} or \t{None} if the queue is empty, as indicated by the declared type \t{Optional[int]}.

The \t{enqueue(v)} function uses \t{cas} to switch \t{buf} from
\t{None} to \t{v} and is atomic since failing \t{cas} operations are
both-movers.
The \t{dequeue()} function use the action \t{r \teq{} buf} to denote
an \emph{unstable read} of \t{buf} that can load any value into the
local variable \t{r}~\cite{DBLP:conf/issta/FlanaganFQ04}.  
Unstable reads can be treated as right-movers since they
commute past steps by other threads.\footnote{Unstable reads are a proof technique that  trades off our ability
to reason about the value stored in \t{r} for the ability to treat the unstable read as a right-mover.
An implementation of unstable read may exhibit any a subset
of the allowed behaviors, including simply performing a conventional read.}
  Consequently, the
\t{dequeue()} function is atomic.  All executions of that function
consist of unstable reads (right-movers) and failed \t{cas} operations
(both-movers) followed by a successful \t{cas} (non-mover).  These
sequences match the reducible pattern $\MR^* [\MN] \ML^*$.  Moreover,
the final \t{cas} ensures that \t{r} is equal to the pre-\t{cas} value
of \t{buf}, which enables mover logic to establish the desired
post-conditions \verb{\result == \old(buf){ and \verb{buf == None{.

\subsection{Lock-Free Stack}
\label{sec:stack}

Figure~\ref{fig:spin} (bottom right) shows a lock-free stack.  This
examples uses immutable lists, where \t{Nil} is the empty list,
$v\t{::}s$ adds $v$ to the front of the the list $s$, and
$\t{head}(s)$ and $\t{tail}(s)$ extract the first element and the rest
of $s$, respectively.\footnote{The duplicated code in this example could be removed in a language with richer control structures such as \t{break} statements.}

The \t{push(v)} function is atomic since it has only one non-mover
operation, namely the successful \t{cas}.  The unstable reads, list
allocations \t{v::t}, and failed \t{cas} operations are both-movers or
right-movers.  Therefore, we can assign \t{push(v)} the following
intuitive post-condition without needing to stabilize under the rely
assumption of a particular caller.
\begin{center}
\begin{tabular}{c}
\begin{lstlisting}[basicstyle=\tt]
ensures head(top) == v
ensures tail(top) == \old(top)
\end{lstlisting}
\end{tabular}
\end{center}

The \t{pop()} function is also atomic due to similar reasoning and
satisfies the following post-condition without the need to stabilize it.
\begin{center}
\begin{tabular}{c}
\begin{lstlisting}[basicstyle=\tt]
ensures head(\old(top)) == \result
ensures tail(\old(top)) == top
\end{lstlisting}
\end{tabular}
\end{center}

\myclearpage
\section{Mover Logic Language}
\label{sec:lang}

\begin{figure}[t!]
  \[
  \begin{array}{l}
  \mbox{\textbf{Syntax}} \\~\\
  \begin{array}{@{~}lrcl}
    (\mbox{\emph{Statements}}) &
    s & ::= & \phantom{|~~}\tskip ~~|~~ \twrong ~~|~~ A ~~|~~ s;s ~~|~~ \sif{\boolcond}{s}{s} \\
      &  &    & |~~ \swhile{\boolcond}{s} ~~|~~ f() ~~|~~ \tyield  \\
    (\mathit{Action}) &        
    A & \subseteq & \Tid \times \Store \times \Store \\
    (\mathit{Thread~Identifier}) & \at, \au & \in & \Tid = \{ 1, 2, \ldots \}\\
    (\mbox{\emph{Conditional Action}}) &            
    \boolcond & ::= & \dotact{A}{A}\\  
    (\mbox{\emph{Variable Declaration}})
    & \fs{var} & ::= & x~var\_spec \\
    & x,y,r,m & \in & \Var \\
    (\mbox{\emph{Function Declaration}})
    & \fs{fn} & ::= & \fs{fn\_spec}~~f()~~\{~ s ~\} \\
    & f & \in & \fs{FunctionName} \\
    (\mbox{\emph{Declaration Table}}) &    
    D & ::= & \overline{\fs{var}~|~\fs{fn}}  \\ & & & \mbox{(\emph{$D$ is left implicit in the semantics for  brevity})}\\
    ~\\
  \multicolumn{4}{@{}l}{\textbf{Semantics}} \\~\\
    (\mathit{Store}) & \sigma & \in & \Var \rightarrow \Value \\
    (\mathit{State}) & \Sigma & ::= & \pstate{\sdots{s_1}{s_n}}{\sigma} \\
    (\mathit{Evaluation~Context}) & E & ::= & \bullet ~|~ E; s 
  \end{array} \\
  ~\\
  \begin{array}{@{\quad}l@{\quad}rcll}
    \multicolumn{3}{l}{\myheadrule{\pstate{s}{\sigma} \rightarrow_{\at} \pstate{s'}{\sigma'}}} \\[2ex]
    \eeval{E-seq}{\at}{\pstate{E[\tskip;s]}{\sigma}}{\pstate{E[s]}{\sigma}}{} \\
    \eeval{E-yield}{\at}{\pstate{E[\tyield]}{\sigma}}{\pstate{E[\tskip]}{\sigma}}{} \\
    \eeval{E-action}{\at}{\pstate{E[A]}{\sigma}}{\pstate{E[\tskip]}{\sigma'}}{\mbox{$\!\!\!\!\!\!\!\!\!\!\!\!\!\!\!\!\!\!\!\!\!\!\!\!\!\!\!\!\!\!\!\!\!\!\!\!\!\!\!\!\!\!\!\!\!\!\!\!\!\!\!\!\!\!\!\!\!\!\!\!\!\!\!\!\!\!\!\!\!\!\!\!$ if $(\at, \sigma,\sigma') \in A$}} \\
    \eeval{E-if}{\at}{\pstate{E[\sif{(\dotact{A_1}{A_2})}{s_1}{s_2}]}{\sigma}}{
        \pstate{E[s_i]}{\sigma'}
    }{\mbox{$\!\!\!\!\!\!\!\!\!\!\!\!\!\!\!\!\!\!\!\!\!\!\!\!\!\!\!\!\!\!\!\!\!\!\!\!\!\!\!\!\!\!\!\!\!\!\!\!\!\!\!\!\!\!\!\!\!\!\!\!\!\!\!\!\!\!\!\!\!\!\!\!$ if $(\at, \sigma,\sigma') \in A_i$, for $i \in 1,2$}} \\
    \eeval{E-while}{\at}{\pstate{E[\swhile{\boolcond}{s}]}{\sigma}}{\pstate{E[\sif{\boolcond}{(s; \swhile{\boolcond}{s})}{\tskip}]}{\sigma}}{} \\
    \eeval{E-call}{\at}{\pstate{E[f()]}{\sigma}}{\pstate{E[s]}{\sigma}}{\mbox{$\!\!\!\!\!\!\!\!\!\!\!\!\!\!\!\!\!\!\!\!\!\!\!\!\!\!\!\!\!\!\!\!\!\!\!\!\!\!\!\!\!\!\!\!\!\!\!\!\!\!\!\!\!\!\!\!\!\!\!\!\!\!\!\!\!\!\!\!\!\!\!\!$ if $\mathit{fn\_spec}~~f()~~\{~ s ~\} \in D$}} \\        
  \end{array}
  ~\\~\\
  \begin{array}{l}
    \myheadrule{\Sigma \rightarrow \Sigma'}\\
  \end{array}\\
  \multicolumn{1}{c}{
    \nrln{E-State}{
      \pstate{s_t}{\sigma} \rightarrow_t \pstate{s'_t}{\sigma'}
    }{
      \pstate{\sdots{\sdots{s_1}{s_t}}{s_n}}{\sigma} \rightarrow \pstate{\sdots{\sdots{s_1}{s'_t}}{s_n}}{\sigma'}
    }
  }
  \end{array}    
  \]
  \caption{\label{fig:sem}Mover Logic Language.}
\end{figure}

\newcommand{\old}[1]{\t{\textbackslash old(#1)}}
\newcommand{\uact}[2]{\langle #1 \rangle_{#2}}

We formalize {\logic} for the idealized language MML (mover logic
language), which we summarize in Figure~\ref{fig:sem}.  
Section~\ref{sub:example} below translates our running example into MLL. 
In MLL, threads manipulate a shared store $\sigma$ that
maps variables to values.  
Variables include
$x$,$y$,$z$, and $m$.  We often use the variable $m$ as a lock, where
$m$ is the thread identifier ($\tid$) of the thread holding the
lock, or 0 if it is not held.

Thread-local variables $r$ are supported by having each thread access
a separate variable $r_{\tid}$ for each thread $\tid$. The language
includes reads and writes to global and local variables, acquires and
releases of locks, local computations, etc.  For generality and
simplicity, we abstract all of these store-manipulation operations as
\emph{actions} $A \subseteq \Tid \times \Store \times \Store$.  Note
that an action may depend on the current thread's identifier.
We write actions as formulae in which $\old{$x$}$ and $x$ to refer to
the values of $x$ in the pre-store and post-store, respectively. We
write $\uact{A}{x}$ to denote an action that only changes $x$:
\[
\uact{A}{x} \defeq \{~ (\tid, \sigma,\sigma') ~~|~~ (\tid, \sigma,\sigma') \in A ~\wedge~
                       \forall y \in \Var \qsep y \neq x \Rightarrow \sigma(y)=\sigma'(y) ~\}
\]
We can then express assignments and locking operations as follows.
Note that \t{acquire($m$)} blocks if the lock is already held,
\emph{i.e.} if $\old{{$m$}} \neq 0$.  We use the notation
$expr[{x}:=\old{{$x$}}]$ to denote $expr$ with all occurrences of ${x}$
replaced by $\old{{$x$}}$.
\[
\begin{array}{rcl}
  \t{acquire($m$)}   & \defeq & \uact{\old{{$m$}} = 0 \wedge {m}=\tid}{{m}} \\
  \t{release($m$)}   & \defeq & \uact{{m}=0}{{m}}   \\
  x ~\t{=}~ expr              & \defeq & \uact{{x} = expr[{x}:=\old{{$x$}}]}{{x}} 
\end{array}
\]
The unstable read $r_\tid ~\teq~ x$ from
Section~\ref{sec:stack} may store any value\footnote{
  In a language with types, this definition can be easily adapted to only store type-correct values into $r_\tid$.
} in the local variable
$r_\tid$:
\[
\begin{array}{rcl}
  r_\tid ~ \teq ~ x  & \defeq & \{~ (\tid, \sigma, \sigma[r_\tid := v]) ~|~ v \in \Value ~\}
\end{array}
\]

Mover Logic Language includes \t{if} and \t{while} statements that
condition execution either on whether a Boolean test is true or on
whether a store-manipulating operation, such as \t{cas}, succeeds.  To
handle these two cases uniformly, we introduce a \emph{conditional
  action} $\boolaction=\dotact{A_1}{A_2}$ where $A_1$ is an action
capturing a true test or successful operation and $A_2$ is an action
capturing a false test or failed operation.  For generality, both
cases may modify the store and both may be feasible on some
pre-states.

We encode any state predicate $B \subseteq \Store$ as the conditional
action $\dotact{\{ (\tid, \sigma, \sigma) ~|~ \sigma \in B \}~}{~\{ (\tid, \sigma,
  \sigma) ~|~ \sigma \not\in B \}}$ that distinguishes the true/false
cases but never modifies the store.  The following illustrates this
encoding for the test $x ~\t{>=}~ 0$.
\[
\begin{array}{rcl}
  x  ~\t{>=}~ 0 & \defeq &   \dotact{\{ (\tid, \sigma, \sigma) ~|~ \sigma(x) \geq 0 \}~}{~\{ (\tid, \sigma, \sigma) ~|~ \sigma(x) < 0 \}}
\end{array}
\]
As a more interesting example, we encode \t{cas} as the following conditional action:
\[
\begin{array}{rcl}
  \t{cas($x$,$v$,$v'$)} & \defeq & \dotact{\uact{\old{{$x$}} = v \wedge {x}=v'}{{x}} ~}{~ I}
\end{array}
\]
where the identity action $I = \{~ (t, \sigma,\sigma) ~~|~~ t \in \Tid \mbox{~and~} \sigma \in
\Store \}$.  This definition permits \t{cas} to non-deterministically
fail from any pre-state, which enables us to treat failed \t{cas}
operations as both movers~\cite{DBLP:journals/pacmpl/FlanaganF20}.

Given $\boolaction = \dotact{A_1}{A_2}$, the if statement
$\sif{\boolaction}{s_1}{s_2}$ may either: 1) evaluate the action $A_1$
and then $s_1$, or 2) evaluate $A_2$ and then $s_2$.  The former is
the ``true'' case and the latter is the ``false'' case, with the
desired behavior regardless of whether $\boolaction$ encodes a
predicate test or a potentially-failing store update.  To prevent the
if statement from blocking, we require $(A_1 \cup A_2)$ to be total on
the state, \emph{i.e.} $\{~ \sigma ~|~ (t, \sigma, \_) \in (A_1 \cup
A_2) ~\} = \State$.

The while statement $\swhile{\boolcond}{s}$ behaves similarly.  It
iterates as long as $\boolaction$ succeeds.  We may need to test
the negation of a conditional action.  The negation of $\boolaction =
\dotact{A_1}{A_2}$, written $!\boolaction$, is simply
$\dotact{A_2}{A_1}$.  
The language includes the statement $\twrong$ to indicate than an
error occurred.  The statement \t{assert B} abbreviates
$\sif{B}{\tskip}{\twrong}$.  The goal of {\logic} is to verify that
programs do not go wrong.

Global variable declarations have the form $x~\fs{var\_spec}$ and are
kept in a global declaration table $D$.  Function declarations have the
form ${fn\_spec}~f()~\{ ~s~ \} $ and are also kept in $D$.
Specifications for globals ($\fs{var\_spec}$) and functions
($\fs{fn\_spec}$) are described in Sections~\ref{sec:effects}
and~\ref{sec:logic}, respectively.  For notational simplicity, $D$ is
left as an implicit argument to the evaluation judgments.
To keep the core language as simple as possible, we elide formal
parameters and return values.  Instead, parameters and return values
are passed in thread-local variables, as described below in
Section~\ref{sub:example}.\footnote{
Extending the language to include function
arguments and results is straightforward, but it adds
notational complexities that are orthogonal to our core contributions.}

In our examples, we include types, curly braces, semicolons, and other
standard syntactic forms to aid readability.

An execution state
\[
\Sigma = \pstate{\sdots{s_1}{s_n}}{\sigma}
\]
consists of sequence of threads $\sdots{s_1}{s_n}$ with a shared store
$\sigma$. The evaluation relation $\Sigma \rightarrow \Sigma'$ is
based on evaluation contexts $E[\ldots]$, which identify the next
statement to be evaluated.
A state $\Sigma = \pstate{\sdots{s_1}{s_n}}{\sigma}$ is \emph{wrong} if any
thread is about to execute \t{wrong}, \emph{i.e.}, if $s_i = E[\twrong]$. 
The semantics demonstrates that \tyield{} annotations have no effect at run
time, but they are used in the mover logic described below.

\myclearpage
\section{Mover Logic Effects and Specifications}
\label{sec:effects}

{\Logic} divides the execution of each thread into reducible code
sequences that are separated by $\tyield$ statements identifying where
thread interference may be observed.  

\subsection{Effects}
\label{sub:effect}

\newcommand{\eg}[0]{\emph{e.g.}}

We use a language of effects to reason about reducible code
sequences separated by {\tyield}s:
\[
\mover \in \mathit{Effect} ~~::=~~ \MY ~|~ \MR ~|~ \ML ~|~ \MB ~|~ \MN ~|~ \ME 
\]
where
\begin{itemize}
\item $\MY$ is the effect of a yield annotation;
\item $\MR$ describes right-mover actions;
\item $\ML$ describes left-mover actions;
\item $\MB$ describes both-mover actions that are both left- and right-movers;
\item $\MN$ describes non-mover actions that are neither left- nor right- movers; and
\item $\ME$ describes erroneous situations, such as the sequential
  composition of two non-mover actions without an intervening \tyield,
  which is not a reducible sequence.
\end{itemize}


Our strategy for verifying that $\tyield$s correctly separate
reducible sequences is based on the DFA~\cite{DBLP:conf/tldi/YiF10} shown below (left). The DFA
captures reducible sequences $\MR^*[\MN]\ML^*$ separated by yields $\MY$, which resets the
DFA to the initial ``pre-commit'' state on the left to start a new
reducible  sequence.  The first left-mover or non-mover
in a reducible sequence is often called the \emph{commit} action and
moves us from the pre-commit to the post-commit phase.

\begin{center}
\begin{tabular}{c@{\qquad}@{\qquad}c}
\includegraphics[width=0.4\textwidth]{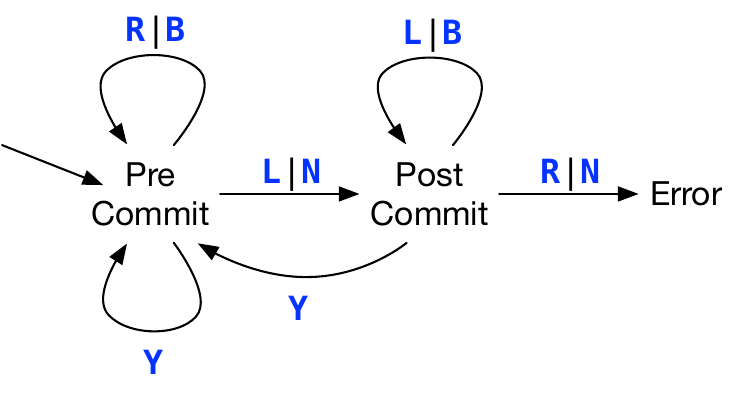} &
\includegraphics[width=0.09\textwidth]{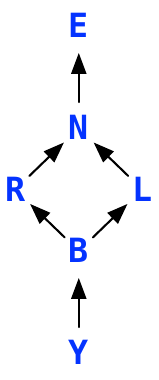}
\end{tabular}
\end{center}
From this DFA, we derive the ordering $\MY \sqsubseteq \MB \sqsubseteq \MR, \ML \sqsubseteq \MN \sqsubseteq \ME$, which is also
shown above~(right).
For example, $\MR \sqsubseteq \MN$, since for any effect sequences
$\alpha$ and $\beta$, if $\alpha\,\MN\,\beta$ is accepted by this DFA,
the $\alpha\,\MR\, \beta$ is also accepted.
We define a standard join operation $\sqcup$ via this ordering.

We
also define sequential composition $\mover_1;\mover_2$ and iterative
closure $\mover^*$, as in~\cite{DBLP:conf/tldi/YiF10}.  For example,
$\MR ; \ML = \MN$ since to show $\alpha\,\MR\,\ML\,\beta$ is accepted by
the DFA it is sufficient to show that $\alpha\,\MN\,\beta$ is accepted.
Conversely, $\MN; \MN = \ME$ (error), since $\alpha\,\MN\,\MN\,\beta$ is
never accepted by this DFA.
\[
\begin{array}{c@{\qquad\qquad}c}
\begin{array}{c|cccccc}
\mover_1;\mover_2 & \MY & \MB & \MR & \ML & \MN & \ME \\
\hline 
\MY & \MY & \MY & \MY & \ML & \ML & \ME \\
\MB & \MY & \MB & \MR & \ML & \MN & \ME \\
\MR & \MR & \MR & \MR & \MN & \MN & \ME \\
\ML & \MY & \ML & \ME & \ML & \ME & \ME \\
\MN & \MR & \MN & \ME & \MN & \ME & \ME \\
\ME & \ME & \ME & \ME & \ME & \ME & \ME \\
\end{array}
&
\begin{array}{c|c}
\mover & \mover^* \\
\hline 
\MY & \MY \\
\MB & \MB \\
\MR & \MR \\
\ML & \ML \\
\MN & \ME \\
\ME & \ME \\
\end{array}
\end{array}
\]

\subsection{Mover Specifications}
\label{sub:specs}

In \logic, the verification of a thread $\tid$ is
performed in the context of a mover specification 
describing how each program action
$A$ starting in the store $\sigma$ commutes with steps of other
threads.
Thus,  mover specifications $M$ have the type
\[
  M: \fs{Action} \times \Tid \times \Store \rightarrow \fs{Effect} \setminus \{ \MY \}
\]
For example, if action $A$ is a local computation that only accesses thread-local variables, we would naturally have
\[
M(A, \tid, \sigma) = \MB
\]
Alternatively, if a global variable \t{x} is protected by a lock
\t{m}, the write action \t{x = $expr$} might have the mover
specification
\[
M(\t{x = $expr$}, \tid, \sigma) =
  \left\{\begin{array}{ll}
  \MB & \mbox{if $\sigma(\t{m}) = \tid$} \\
  \ME & \mbox{otherwise}
  \end{array}\right.
\]
indicating that the write is a both-mover only if thread $\tid$ holds
lock \t{m}.  Otherwise, it is an error. We assume that $expr$
only accesses local variables, and that $M(A,\tid,\sigma)$ is never
$\MY$ since actions do not yield.  

We write mover specifications in the source code using the following
notation, which is inspired by earlier reduction-based
verifiers~\cite{DBLP:conf/cav/HawblitzelPQT15,DBLP:journals/pacmpl/FlanaganF20,DBLP:journals/toplas/FlanaganFLQ08}:
\[
  \begin{array}{rcl}
      var\_spec & ::= & var\_clause^* \\
      var\_clause   & ::= & \t{read}~\mover~\t{if}~P ~|~ \t{write}~\mover~\t{if}~P
  \end{array}   
\]
where $P\subseteq \Tid \times \Store \times \Store$ is a two-store
predicate describing the pre-store and post-store of the access to $x$
in question.  Further, $P$ can depend on the current thread identifier
$\tid$.  Similar to actions, we write these predicates as formulae in
which $\old{$y$}$ and $y$ to refer to the values of $y$ in the
pre-store and post-store, respectively.
To determine the mover effect of a variable access, we evaluate the
specification clauses in order and take the effect of the first case
where the condition $P$ is satisfied.  If no clauses apply, the access
has the error effect $\ME$.
More formally, given the specification for a variable \t{x} in
the source code, we collect the sequence of clauses for reads and
writes separately and then create the mover specification $M$ for
\t{x} as follows:
\[
  \begin{array}{rcl}
    \left[\!\!\!\left[
    \begin{array}{c}
    \t{read} ~ \mover_1 ~ \t{if} ~ P_1 \\
    \vdots \\
    \t{read} ~ \mover_n ~ \t{if} ~ P_n 
  \end{array}
  \right]\!\!\!\right]
  & \Longrightarrow &
  M(\t{r}_\tid = \t{x}, \tid, \sigma) =
    \left\{\begin{array}{ll}
      \mover_1 & \mbox{if $P_1(\tid, \sigma,\sigma)$} \\
      \multicolumn{1}{c}{\vdots} &       \multicolumn{1}{c}{\vdots} \\
      \mover_n & \mbox{if $P_n(\tid, \sigma,\sigma)$} \\
      \ME & \mbox{otherwise}
  \end{array}\right. 
  \\~\\
  \left[\!\!\!\left[
    \begin{array}{c}
    \t{write} ~ \mover_1 ~ \t{if} ~ P_1 \\
    \vdots \\
    \t{write} ~ \mover_n ~ \t{if} ~ P_n 
  \end{array}
  \right]\!\!\!\right]
  & \Longrightarrow &
  M(\t{x} = expr, \tid, \sigma) =
    \left\{\begin{array}{ll}
      \mover_1 & \mbox{if $P_1(\tid, \sigma,\sigma[\t{x}:=\sigma(expr)])$} \\
      \multicolumn{1}{c}{\vdots} &       \multicolumn{1}{c}{\vdots} \\
      \mover_n & \mbox{if $P_n(\tid, \sigma,\sigma[\t{x}:=\sigma(expr)])$} \\
      \ME & \mbox{otherwise}
  \end{array}\right.  
\end{array}
\]
where $\t{r}_{tid}$ is a local variable, $expr$ only accesses thread-local variables, $\sigma(expr)$ is the result of evaluating $expr$ in the store $\sigma$, and the cases for $M$ are evaluated in the order listed.

The declaration for a global variable \t{x} protected by a lock \t{m} is thus written as
\begin{center}
  \begin{tabular}{c}
  \begin{lstlisting}[basicstyle=\tt]
    int x  read  both-mover if m == tid
           write both-mover if m == tid
  \end{lstlisting}
  \end{tabular}
\end{center}
where \inlinecode{both-mover} is syntactic sugar for the effect
$\MB$.  (Similarly, we use \inlinecode{left-mover} for $\ML$, and so
on.)  In our examples, 
we
abbreviate these identical read and write cases as follows.
\begin{center}
  \begin{tabular}{c}
  \begin{lstlisting}[basicstyle=\tt]
    int x  both-mover if m == tid    $~~$         
  \end{lstlisting}
  \end{tabular}
\end{center}

\begin{figure}[t!]
  \centering
  \includegraphics[width=\textwidth]{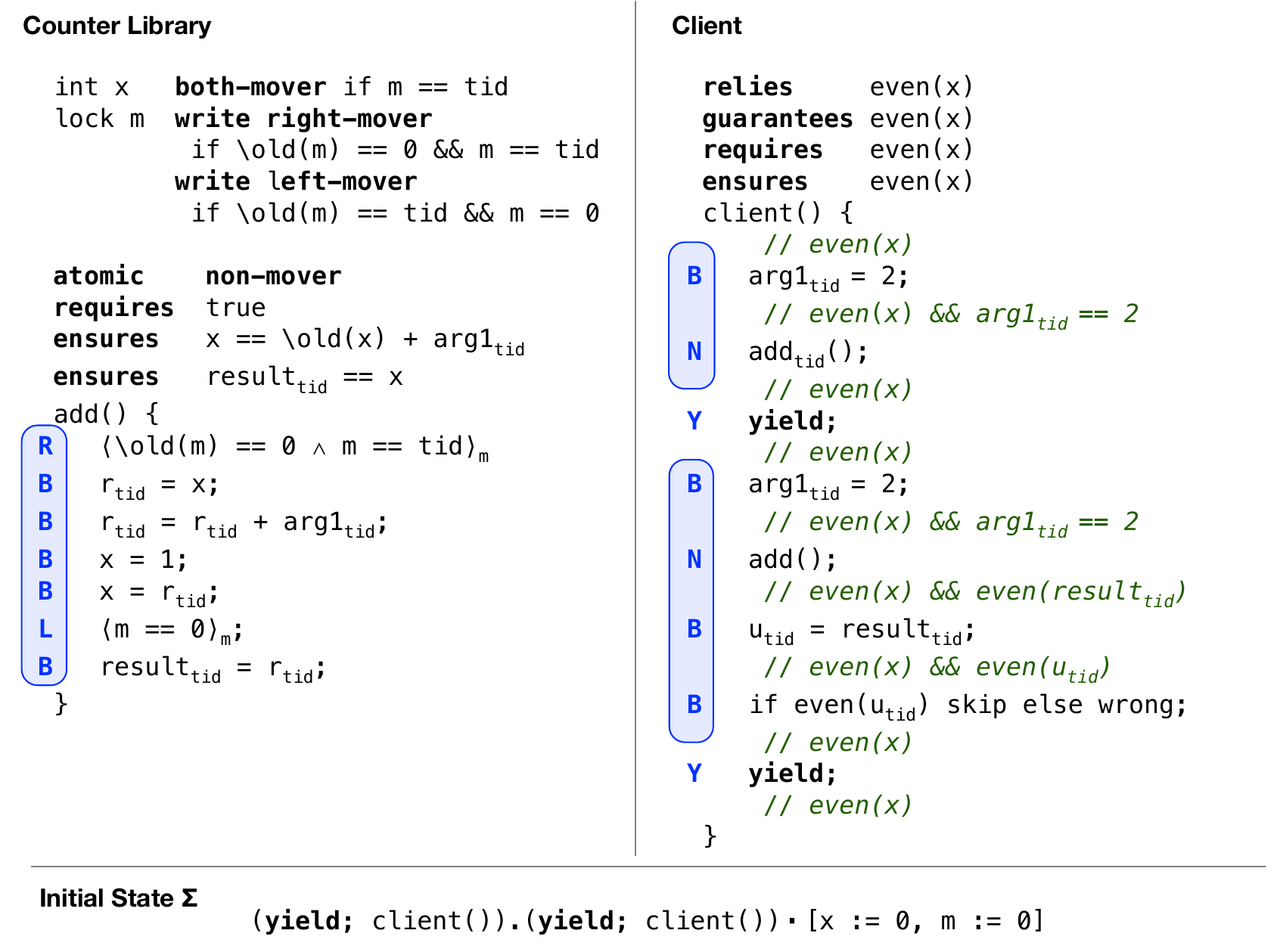}
  \caption{\label{fig:revisit}The example from
    Figure~\ref{fig:set-inv}~(right) expressed in Mover Logic
    Language.}
\end{figure}

\myclearpage
\subsection{Motivating Example, Revisited}
\label{sub:example}

Figure~\ref{fig:revisit} expresses our motivating example from
Figure~\ref{fig:set-inv}~(right) in our Mover Logic Language.
As mentioned earlier, an access to a thread-local variable $r$
actually accesses a (global) variable $r_{\tid}$ that is reserved for
use by thread $\tid$.
We use thread-local variables to encode function arguments and
results.  The fork statements are converted into parallel threads in
the initial state $\Sigma$.
%
We insert a \t{yield} at the start of each thread in $\Sigma$ so that
the initial state is well-formed under the non-preemptive semantics we
introduce in our formal development.

Given this mover specification, mover logic successfully verifies this
code. Figure~\ref{fig:revisit} also demonstrates the reasoning carried
out by mover logic.  The left margin shows the effect of
each action and groups those effects into reducible sequences.  The
\t{add()} function is a single reducible sequence, ensuring that
we may treat it as atomic.  The \t{client()} function consists of
multiple reducible sequences separated by $\tyield$s.  We also show
invariants demonstrating that \t{client()} is correct in comments
at each program point.\footnote{In this example, the rely assumption
  \t{even(x)} is sufficient for reasoning about \tyield{} points.  In
  code where live ranges for local variables span yield points, we
  would add to the rely assumptions the requirement that one thread
  does not change another thread's local variables.}

\myclearpage
\subsection{Additional Mover Specification Examples}
\label{sub:more-ex}

Figure~\ref{fig:set-inv}~(right) showed how mover specifications can
capture the synchronization/commuting behavior of lock acquires,
lock releases, and lock-protected variable accesses.
Our mover specifications are inspired by the \textsc{Anchor} verifier,
which used mover specifications to capture many synchronization
idioms~\cite{DBLP:journals/pacmpl/FlanaganF20,anchor-site}.\footnote{Our
syntax for mover specifications is a syntactic variant of the
\textsc{Anchor} syntax.  In essence, our specifications are sequential
\emph{var\_clauses}, whereas \textsc{Anchor} combines these clauses
into a single binary decision tree using the syntax \emph{bool\_expr}
? \emph{mover\_spec} : \emph{mover\_spec}.}

To illustrate how mover specifications capture more complex
synchronization disciplines, suppose the variable \t{y} is
\emph{write-protected} by a lock \t{m}.  That is, lock \t{m} must be
held for all writes to \t{y} but not necessarily held for reads.
Consequently, \t{y} should be declared volatile if the code is run under a weak memory model.
Writes to \t{y} are non-movers (due to concurrent reads);
lock-protected reads are both-movers (because there can be no
concurrent writes); and reads without holding the lock are non-movers
(due to concurrent writes).  Mover specifications capture this
synchronization discipline concisely as follows, where the last clause
applies only when \t{m} is not \t{tid}.
\begin{center}
  \begin{tabular}{c}
    \begin{lstlisting}[basicstyle=\tt\footnotesize]
int y write non-mover  if m == tid
      read  both-mover if m == tid
      read  non-mover
    \end{lstlisting}
  \end{tabular}
\end{center}

The \textsc{FastTrack} dynamic race
detector~\cite{DBLP:journals/cacm/FlanaganF10,DBLP:conf/ppopp/WilcoxFF18}
uses a combination of lock-protected and write-protected disciplines
to synchronize accesses to some array pointers.  We illustrate that
discipline for an array pointer \t{vc}: initially, a \t{flag} is false
and the pointer \t{vc} is guarded by \t{lock}; when \t{flag}
becomes true, \t{vc} becomes write-guarded by \t{lock}.  The mover
specification for this discipline is captured by the first four lines
in the specification for \t{vc}:
  \begin{center}
    \begin{tabular}{c}
    \begin{lstlisting}[basicstyle=\tt]
      int vc[]        both-mover if !flag && lock == tid
                write non-mover  if  flag && lock == tid
                read  both-mover if  flag && lock == tid
                read  non-mover  if  flag
                [i]        both-mover if !flag && lock == tid
                [i] read   both-mover if  flag && (lock == tid || tid == i)
                [i] write  both-mover if  flag && (lock == tid && tid == i)
    \end{lstlisting}
    \end{tabular}
  \end{center}
This idiom enables the algorithm to avoid using a lock to protected
all accesses to \t{vc} but still replace \t{vc} with a larger array
when necessary.  The last three lines capture the synchronization
discipline for accessing the array entry \t{vc[i]}, where we use the
extended notation ``$ \t{[i]}~\mathit{var\_clause} $''
to describe the synchronization cases for actions that access
\t{vc[i]}.  That entry is also initially guarded by \t{lock} when
\t{flag} is false; when \t{flag} becomes true, the entry \t{vc[i]} can
only be written by thread \t{i} while holding \t{lock}, read by any
thread while holding the lock, or read by thread \t{i} without holding
the lock.  These reads and writes are all both-movers.  These rules
prevent all conflicting reads and writes, and thus all accesses to
\t{vc[i]} are both-movers under this synchronization discipline.

As a final example, consider a concurrent hashtable consisting of a
\t{table} array and a \t{locks} array, which has length \t{N}.  The
entry \t{table[i]} is protected by \t{locks[i \% N]}.  The \t{table}
reference itself may change when, for example, \t{table} is replaced
with a larger array.  To ensure such changes are done without
interference, a write to \t{table} is permitted only when a thread
holds \emph{all} locks.  In contrast, \t{table} can be read by a
thread holding \emph{any} lock.  All such reads and writes are
both-movers, as captured by the following mover
specification:
  \begin{center}
    \begin{tabular}{c}
    \begin{lstlisting}[basicstyle=\tt]
    Entry table[] write both-mover if $\forall \t{i} \in [0,\t{N})\qsep$locks[i] == tid
                  read  both-mover if $\exists \t{i} \in [0,\t{N})\qsep$locks[i] == tid
                  [i]   both-mover if locks[i % N] == tid
    \end{lstlisting}
    \end{tabular}
  \end{center}

As illustrated in the previous two examples from the \textsc{Anchor}
verifier~\cite{DBLP:journals/pacmpl/FlanaganF20}, mover specifications
can naturally capture synchronization disciplines that vary with the current
program state.  

A final example comes from the common iterative
parallel algorithm pattern in which a synchronization barrier is used
to divide the computation into a series of phases. In the even phases,
the main thread (with \t{tid} = 0) updates shared data structures, and in
odd phases, worker threads concurrently read data from those
structures, as specified below.
  \begin{center}
    \begin{tabular}{c}
    \begin{lstlisting}[basicstyle=\tt]
    int z  read both-mover if phase % 2 == 1
                both-mover if phase % 2 == 0 && tid == 0
    \end{lstlisting}
    \end{tabular}
  \end{center}

\begin{figure}[t!]
  \[
  \begin{array}{@{}l}
    {\mbox{\bf One-Store and Two-Store Predicates and Supporting Definitions}}\\~\\
    \multicolumn{1}{c}{
    \begin{array}{@{}rcl}
      S,T & \subseteq & \Tid \times \Store \\
      R,G,P,Q,A & \subseteq & \Tid \times \Store \times \Store \\
    \end{array}} \\
    \begin{array}{@{}l@{}l@{}}
     \begin{array}{@{}r@{~~}c@{~~}l}    
      \two{S}       & \defeq & \{~ (\at, \sigma,\sigma) ~|~ (\at, \sigma) \in S ~\} \\
      \post{P}      & \defeq & \{~ (\at, \sigma ) ~|~ (\at, \_, \sigma) \in P ~\} \\
      I   & \defeq & \{~ (\at, \sigma, \sigma) ~|~ \at \in \Tid, \sigma \in \Store ~\} \\
     \end{array} &
     \begin{array}{r@{~~}c@{~~}l@{}}    
      P;A     & \defeq & \left\{ (\at, \sigma,\sigma'') ~\middle|~
               \begin{array}{@{}l@{}}
                 (\at, \sigma,\sigma') \in P \mbox{~and~}  \\
                 (\at, \sigma',\sigma'') \in A
               \end{array}
               \right\} \\
      \yield{P}{R}  & \defeq & \left\{ (\at, \sigma', \sigma') ~\middle|~ \begin{array}{@{}l@{}}
               (\at, \_,\sigma) \in P \mbox{~and~} \\
               (\at, \sigma, \sigma') \in R^*
               \end{array}
               \right\}         
     \end{array}
    \end{array}     
      \\~\\
      \multicolumn{1}{@{}l}{\mbox{\bf Mover Logic Proof Rules}}\\~\\    
  \begin{array}{@{}l@{~}l}
    \myheadrule{\mimp{R}{G}{s}{P}{Q}{\mover}}
    \\~\\
    \nrln{M-action}{
      M(A,P) \sqsubseteq \mover \\
      e \sqsubseteq \ML \Rightarrow \mbox{$A$ is total}
    }{
      \mimp{R}{G}{A}{P}{(P;A)}{\mover}
    }
    &
    \nrln{M-seq}{
      \mimp{R}{G}{s_1}{~P~}{Q_1}{\mover_1} \\
      \mimp{R}{G}{s_2}{Q_1}{Q_2}{\mover_2} \\            
    }{
      \mimp{R}{G}{s_1;s_2}{P}{Q_2}{(\mover_1;\mover_2)}
    }
    \\~\\
    \nrln{M-if}{
      \mimp{R}{G}{s_1}{P;A_1}{Q}{\mover_1} \\
      \mimp{R}{G}{s_2}{P;A_2}{Q}{\mover_2} \\
      (M(A_1,P);\mover_1) \sqcup (M(A_2,P);\mover_2) \sqsubseteq \mover
    }{
      \mimp{R}{G}{\sif{(\dotact{A_1}{A_2})}{s_1}{s_2}}{P}{Q}{\mover}
    }
    &
    \nrln{M-while}{
      \mimp{R}{G}{s}{P;A_1}{P}{\mover_1} \\
      M(A_1,P);\mover_1 \sqsubseteq \MR \\
      (M(A_1,P);\mover_1)^*; M(A_2,P) \sqsubseteq \mover\\
      \mover \not\sqsubseteq \ML
    }{
      \mimp{R}{G}{\swhile{(\dotact{A_1}{A_2})}{s}}{P}{P;A_2}{e}
    }
    \\~\\
    \nrln{M-skip}{    }{      \mimp{R}{G}{\tskip}{P}{P}{\MB}    }
    &
    \nrln{M-wrong}{    }{      \mimp{R}{G}{\twrong}{\emptyset}{\emptyset}{\MB}    }
    \\~\\
    \nrln{M-conseq}{
    \begin{array}{@{}r@{~}c@{~}l}
      P &\Rightarrow& P_1 \\ Q_1 &\Rightarrow& Q 
    \end{array} 
    \begin{array}{r@{~}c@{~}l}
      R &\Rightarrow& R_1 \\ G_1 &\Rightarrow& G 
    \end{array} 
    \begin{array}{r@{~}c@{~}l@{}}
       ~\\ \mover_1 &\sqsubseteq& \mover 
     \end{array} \\
      \mimp{R_1}{G_1}{s}{P_1}{Q_1}{\mover_1} 
    }{
      \mimp{R}{G}{s}{P}{Q}{\mover}
    }
    &
    \nrln{M-yield}{~\\
      P \Rightarrow G \\
      Q = \yield{P}{R} 
    }{
      \mimp{R}{G}{\tyield}{P}{Q}{\MY}
    }
  \end{array}
  \end{array}
  \]
    \caption{\label{fig:mover-logic} {\Logic} proof rules and supporting definitions.}
\end{figure}

\begin{figure}[p!]
  \[
  \begin{array}{lc}
    \mbox{\bf Function Specification Syntax} \\~\\
    \multicolumn{2}{c}{\begin{array}{rcr@{\ \ }l}
  \fs{fn\_spec} & ::= & \batomic~\mover & \brequires~S~\bensures~Q \\
            &  |  & \brelies~R~\bguarantees~G & \brequires~S~\bensures~T
    \end{array}} \\~\\
    \mbox{\bf Proof Rules for Function Definitions and Calls} \\~\\    
    \myheadrule{\vdash \fs{fn}} \\
    \nrln{M-def-atomic}{
      \mbox{$f()$ is not (directly or indirectly) recursive}\\
      \mimp{\emptyset}{\emptyset}{s}{\two{S}}{Q}{\mover} 
    }{
      \begin{array}{l@{~}l}
        \vdash & {\bf atomic}~\mover \\
               & {\bf requires}~S~{\bf ensures}~Q\ \ \ f()~\{~s~\}
      \end{array}      
    }
    &
    \includegraphics[width=0.4\textwidth,valign=M]{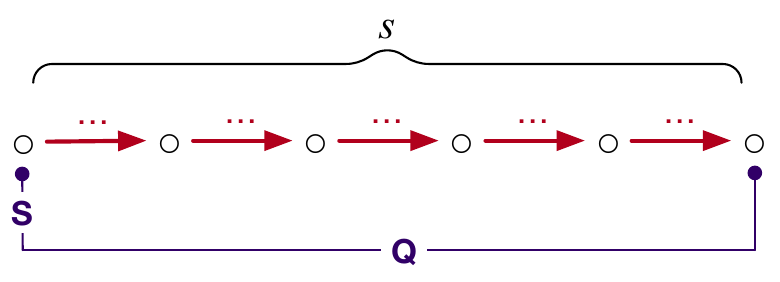}
    \\~\\        
    \nrln{M-def-non-atomic}{
      \mimp{R}{G}{s}{\two{S}}{\two{T}}{\MR} \qquad G \neq \emptyset \\ 
    }{
      \begin{array}{l@{~}l}
        \vdash & {\bf relies}~R~{\bf guarantees}~G \\
               & {\bf requires}~S~{\bf ensures}~T\ \ \ f()~\{~s~\}
      \end{array}      
    }
    &
    \includegraphics[width=0.4\textwidth,valign=M]{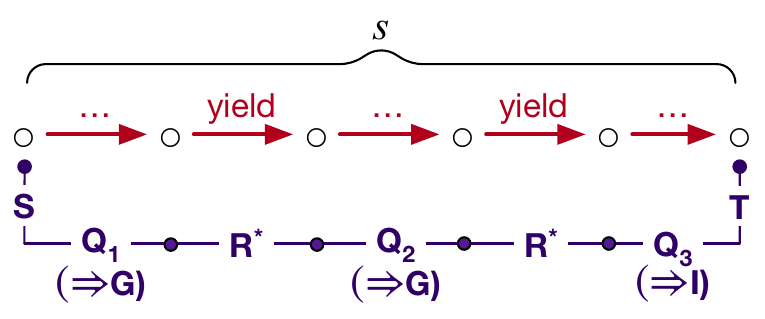}    
    \\~\\
    \myheadrule{\mimp{R}{G}{s}{P}{Q}{\mover}} \\
    \nrln{M-call-atomic}{
      \begin{array}{l}
         {\bf atomic}~\mover\\
         {\bf requires}~S~{\bf ensures}~Q\ \ \ f()~\{~s~\} \in D
      \end{array}      \\
      \post{P} \Rightarrow S \\
    }{
      \mimp{R}{G}{f()}{P}{(P;Q)}{\mover}
    }
    &
    \includegraphics[width=0.4\textwidth,valign=M]{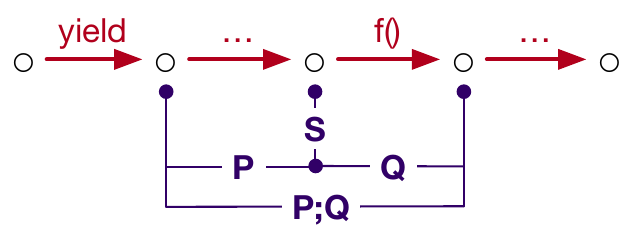}
    \\~\\
    \nrln{M-call-non-atomic}{
      \begin{array}{l}
         {\bf relies}~R~{\bf guarantees}~G \\
         {\bf requires}~S~{\bf ensures}~T\ \ \ f()~\{~s~\} \in D
      \end{array}      \\
    }{
      \mimp{R}{G}{f()}{\two{S}}{\two{T}}{\MR}
    }
    &
    \includegraphics[width=0.4\textwidth,valign=M]{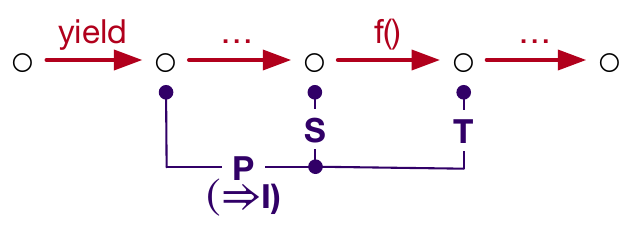}
    \\~\\
    \mbox{\bf Verification of States} \\~\\
    \myheadrule{\vdash \Sigma} \\    
    \multicolumn{2}{c}{
    \nrln{M-state}{
      \forall \fs{fn} \in D\qsep \vdash \fs{fn} \qquad 
      M \mbox{~is valid} \qquad
      I \Rightarrow G \\      
      \forall \at \in \Tid\qsep
      \left[
        \begin{array}{l}
          \mimp{R}{G}{s_\at}{P_\at}{Q_\at}{\mover_\at} \mbox{~and~} \mover_\at \neq \ME \mbox{~and~} Q_\at \Rightarrow G \\
          \mbox{and $s_\at$ is yielding} \mbox{~and~} (\at, \sigma, \sigma) \in P_\at \\
        \end{array}
        \right]\\
      \forall t,u \in \Tid \qsep t \neq u \Rightarrow \left( G[\tid := \at] \Rightarrow R[\tid := \au] \right) \\
    }{
      \vdash \pstate{\sdots{s_1}{s_n}}{\sigma}
    }
    }
  \end{array}
  \]
  \caption{\label{fig:rules-fn} {\Logic} proof rules for function definition, calls, and run-time states.}
\end{figure}


\myclearpage
\section{{\LOGIC}}
\label{sec:logic}

In this section, we show the proof rules for how mover logic handles statements
(Section~\ref{sub:logic}); function
definitions, calls, and specifications
(Sections~\ref{sub:atomic}--\ref{sub:non-atomic}); and run-time states~(Section~\ref{sub:states}).

\subsection{{\LOGIC}}
\label{sub:logic}

Mover logic is defined via the judgments in
Figures~\ref{fig:mover-logic}~and~\ref{fig:rules-fn}.  The main judgment
\[
  \mimp{R}{G}{s}{P}{Q}{\mover}
\]
verifies that, when starting from a store satisfying the precondition
$P$, the statement $s$ terminates only in stores satisfying the
postcondition $Q$ (\emph{i.e.} partial correctness).  In addition, the
judgment uses the mover specification $M$ to verify that $s$ consists
of reducible sequences separated by {\tyield}s.  At each yield point,
the rely assumption $R \subseteq \Tid \times \Store \times \Store$ is used to
model potential interference from other threads.  Conversely, the
thread guarantee $G \subseteq \Tid \times \Store \times \Store$ summarizes the
behavior of each reducible code sequence between two yield points in
$s$.  The effect $\mover$ summarizes how $s$ commutes with steps of
other threads.

In the rules, the precondition $P$ can refer to the value of
variable $\t{x}$ in the initial store $\sigma_0$ of the current
reducible code sequence via the notation $\old{\t{x}}$. Thus $P$ is
a two-store relation $P \subseteq \Tid \times \Store \times \Store$
relating that initial store $\sigma_0$ to the pre-store $\sigma$ for
the execution of $s$.  We show that requirement visually in the
following trace, where $(\tid, \sigma_0, \sigma) \in P$.
\begin{center}
  \includegraphics[width=0.6\textwidth]{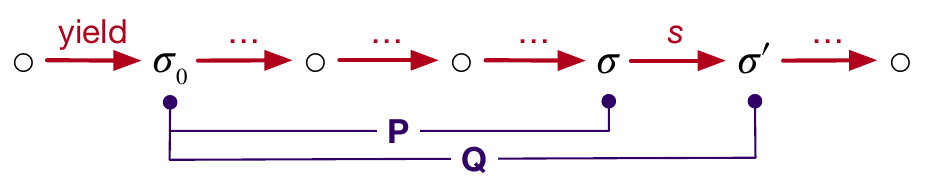}
\end{center}
%
The two-store postcondition $Q \subseteq \Tid \times \Store \times
\Store$ relates $\sigma_0$ to the post-store $\sigma'$ of $s$.
%


Many of the mover logic rules are extensions of Hoare logic
incorporating reduction effects.  For example, the rule
\relname{M-seq} states that a sequential composition $(s_1;s_2)$
commutes as $\mover_1;\mover_2$, the sequential composition of the
effects of its sub-statements, and that the precondition and
postcondition are related as follows:
\begin{center}
  \includegraphics[width=0.5\textwidth]{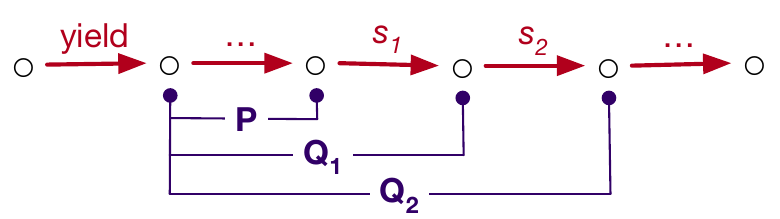}
\end{center}
The rule
\relname{M-skip} indicates that $\tskip$ has no effect, so its
precondition and postcondition are identical.
The rule \relname{M-wrong} verifies that $\twrong$ is never executed
via the unsatisfiable precondition $\emptyset$.   That is, this rule
rejects any program that may execute $\twrong$ from any state.
The
rule~\relname{M-action} bounds the effect of action $A$ from
states $\sigma$ satisfying the current precondition $P$.  That rule
uses the function
\[
 M(A,P) ~~\defeq~~ \bigsqcup_{ (\at,\_,\sigma) \in P} M(A,\at,\sigma)
\]
(note that we are overloading $M$ here)
to compute the join of $A$'s mover behavior over all such states, and
requires only that the ascribed effect $\mover$ over-approximate it,
\emph{i.e.} $M(A,P) \sqsubseteq \mover$.  Ascribing a larger effect is
always sound, since a greater effect is a weaker commutativity claim;
this built-in weakening subsumes an application of \relname{M-conseq}
to the effect and lets each action carry an effect compatible with its
context\footnote{This rule presented in the formal development of the ECOOP 2024
version~\cite{DBLP:conf/ecoop/FlanaganF24} incorrectly stated the effect antecedent of \relname{M-action} as an equality, which was too restrictive to prove the Prefix lemma
(Lemma~\ref{lem:prefix}), on which the proof of Preservation
(Theorem~\ref{thm:pres}) relies.  Similar changes were made to the rules \relname{M-if} and \relname{M-while} in this version of the paper.}.  The rules \relname{M-if} and \relname{M-while} adopt the same
upper-bound form for their effect antecedents, which is essential for the
Prefix lemma (Lemma~\ref{lem:prefix}) below: strengthening a precondition
shrinks $M(A,P)$, so an exact effect computation could not be re-established
under a stronger precondition, whereas an upper bound transfers by
monotonicity.
The postcondition of $A$ is then the precondition $P$
sequentially composed with the action $A$, \emph{i.e.} $P;A$.
A key technical requirement of the reduction theorem is that once an
atomic block $\MR^*[\MN]\ML^*$ enters its post-commit (or left-mover
part), then it must terminate. It cannot block or
diverge~\cite{DBLP:conf/tldi/FlanaganQ03}.\footnote{ To motivate this
requirement consider the program \t{(x = 1; while (true) skip; yield)
  || (assert x != 1)}.  This program can go wrong because the
first thread writes 1 to \t{x}.  However, the reducible block
containing that write never terminates after performing that write,
and that write is not included in the thread guarantee $G$.  Thus, we
require that once a reducible block commits, it must terminate.}
Hence, we require that $A$ is total if it is a left-mover.  We 
place similar restrictions on loops.

The rule \relname{M-if} requires both the true case ($A_1; s_1$) and
the false case ($A_2; s_2$) to have the same post-condition $Q$.  The
effect $e$ bounds the maximal effect of executing either $A_1$ followed by
$s_1$ or $A_2$ followed by $s_2$.
The rule \relname{M-while} for $\swhile{\dotact{A_1}{A_2}}{s}$ checks
that a successful test followed by the body preserves precondition
$P$, which functions as a loop invariant.  The postcondition of the
loop is the postcondition of $A_2$ given the precondition $P$.  The
effect $e$ bounds the iterative closure of the effect of one
iteration sequentially composed with the effect of the
loop-terminating test $A_2$.  The antecedent
$M(A_1,P);\mover_1 \sqsubseteq \MR$ requires each iteration to behave as a
right-mover, so an iteration cannot commit and keep looping (an iteration may
commit only if it also yields, resetting to a new reducible block); together
with $\mover \not\sqsubseteq \ML$, which prevents the loop from being placed
after the commit point of an enclosing reducible block, this ensures a loop
never runs post-commit, where the termination requirement discussed above
would apply.

Consider the loop in \t{spin\_lock()} in Figure~\ref{fig:spin}.  The
test \t{!cas(l,0,tid)} is the conditional action
$\dotact{I}{\uact{\old{{l}} = 0 \wedge \t{l} =
    \t{tid}}{\t{l}}}$ and the loop body is \t{skip}.  Since $P;I =
P$, rule~\relname{M-skip} concludes that
$\mimp{R}{G}{\tskip}{P}{P}{\MB}$.  Further, $M(I,P) = \MB$, because
that action accesses no global variables, and the specification for
\t{l} indicates that $M(\uact{\old{{l}} = 0 \wedge \t{l} =
  \t{tid}}{\t{l}}, P) = \MR$.  Thus, $e = (\MB;\MB)^*;\MR = \MR$.
Also, the postcondition $P;A_2$ for the loop simplifies to the
expected $P[\t{l}:=\t{tid}]$.
To ensure the left-mover termination requirement,
rule~\relname{M-while} requires that $e \not\sqsubseteq\ML$.  That is,
the post-commit part of a reducible sequence cannot contain loops.

The rule \relname{M-yield} for $\tyield$ first checks that the thread
guarantee $G$ includes all possible behaviors $P$ of the reducible
sequence preceding the $\tyield$ via the antecedent $P \Rightarrow G$.
The reducible sequence following the $\tyield$ starts with
postcondition
$
Q = \yield{P}{R}
$
which incorporates repeated thread interference from other threads via
the iterated rely assumption $R^*$ and then resets each $\old{x}$
value to be the current value of $\t{x}$ at the start of the new
reducible sequence.

The rule \relname{M-conseq} extends the consequence rule of RG
logic to reduction effects.

\subsection{Atomic Functions}
\label{sub:atomic}

{\Logic} supports both atomic and non-atomic functions. An atomic
function is one whose code body is reducible (\emph{i.e.}, no \t{yield}
statements) and has the following form:
\[
      \begin{array}{l}
         {\bf atomic}~\mover \\
         {\bf requires}~S~{\bf ensures}~Q\ \ \ f()~\{~s~\}
      \end{array}
\]
(We elide $\mover$ in the surface
syntax when it is $\MN$, as in Figure~\ref{fig:set-inv}~(right)).      
The precondition $S \subseteq \Tid \times \Store$ describes valid
initial stores for the function call and must be established by the
caller. The post condition $Q \subseteq \Tid \times \Store \times
\Store$ describes possible final stores, and it may refer to values of
variables on function entry using the \t{\old{x}} notation.
%
%
Since $s$ is atomic and \tyield-free, we elide the rely
and guarantee components from atomic function  specifications.  We require
atomic functions to be non-recursive to facilitate the ``left-mover
terminates'' requirement mentioned above.

To ensure that the function body $s$ conforms to the function's
specification, rule \relname{M-def-atomic} in
Figure~\ref{fig:rules-fn} first converts $S$ into the two-store
precondition $\two{S}$ (in which $\verb{\old(x){ = \t{x}$ for all
variables \t{x}) and then verifies the function body $s$ with respect
to that precondition.  We use the guarantee $\emptyset$ to enforce
that $s$ is indeed yield-free.  (Rule~\relname{M-yield} will always
fail if $G$ is $\emptyset$, provided that the \tyield{} is actually
reachable, \emph{i.e.} if $P \neq \emptyset$).

The rule \relname{M-call-atomic} for a corresponding call to $f()$
retrieves the above specification from the declaration table $D$ and then
ensures that the precondition $P$ at the call site implies the
callee's precondition $S$.  That rule uses $\post{P}$ to first convert $P$ into a one-state predicate.  
The postcondition $(P;Q)$ combines the
call precondition $P$ with the two-store postcondition $Q$ of the
callee, as illustrated in the trace to the right of the rule.

\subsection{Non-Atomic Functions}
\label{sub:non-atomic}

Non-atomic function definitions have the following form:
\[
      \begin{array}{l}
         {\bf requires}~R~{\bf guarantees}~G\\
         {\bf requires}~S~{\bf ensures}~T\ \ \ f()~\{~s~\}
      \end{array}
\]
We include thread rely $R$ and guarantee $G$ components in these
function specifications since non-atomic function may include
\tyield{} points where thread interference may occur.  For simplicity, we
require that non-atomic function calls and returns happen at the start
of a reducible sequence.  Consequently, the precondition $S \subseteq \Tid \times
\Store$ and postcondition $T \subseteq  \Tid \times \Store$ are both one-store
predicates since there is no need to summarize the preceding reducible
sequence.

The rule \relname{M-def-non-atomic} checks that the function body $s$
runs from the precondition $\two{S}$, possibly via multiple reducible
sequences separated by {\tyield}s, to terminate after a final $\tyield$ s
in a store satisfying $T$. Those requirements are enforced
by using $\two{T}$ as the postcondition for $s$.  Further,
the body $s$ should end in a yield, which from the definition of $\mover_1;\mover_2$ entails that the 
effect of $s$ is at most  $\MR$.
At a
call site, the rule \relname{M-call-non-atomic}
requires that the current reducible sequence is trivial/empty and
meets the function's one-store precondition $S$ by requiring the
precondition $\two{S}$ prior to the call.  The
rule also converts the function's one-store
postcondition $T$ to the two-store predicate $\two{T}$.

\subsection{Verifying States}
\label{sub:states}

We now define the verification judgment $\vdash \Sigma$ to verify program states $\Sigma =
\pstate{\sdots{s_1}{s_n}}{\sigma}$.  The rule~\relname{M-state} for
this judgment in Figure~\ref{fig:rules-fn} ensures that:
\begin{itemize}
\item
  each thread $s_t$ verifies from a precondition $P_t$ that includes the initial store
    $\sigma$;
\item
  any pending behavior $Q_t$ at thread termination is
    published to $G$;
  \item
    the thread guarantee $G$ is reflexive;
  \item
    the guarantee of each thread is contained in
    the rely assumption of every other thread; 
  \item
    each function definition in the global table $D$ is verifiable; and 
  \item
    that all threads start with a $\tyield$ statement (to simplify the correctness proof).
\end{itemize}

A mover specification $M$ makes claims about how steps of one thread commute with respect to steps of other threads, and mover logic needs to ensure that those claims are correct. Specifically, we define a mover specification to be \emph{valid} if:
\begin{enumerate}
\item Right-moving actions can be moved later in a trace without changing the final store. 
\item Left-moving actions can be moved earlier in a trace without changing the final store.     
\item An action by one thread cannot change the effect of an action in another thread.
\item An action by one thread cannot cause a left-moving action in
  another thread to block.  
\end{enumerate}  
We formalize these validity requirements as follows:

\begin{definition}[Validity]
  $M$ is \emph{valid} if
  the following four conditions hold for all threads $\at \neq \au$ and $A_1, A_2, \sigma, \sigma'$:
  \begin{center}
  \begin{tabular}{l@{\qquad}c}
    \begin{tabular}{ll}
      ~\\
    (1)
    & if $M(A_1,\at,\sigma) \sqsubseteq \MR$ and $(\at, \sigma, \sigma') \in A_1$ and \\
    & $\phantom{\mbox{if}}$ $M(A_2, \au, \sigma') \sqsubseteq \MN$ and $(\au, \sigma', \sigma'') \in A_2$,\\
    & then there exists $\sigma'''$ such that \\
      & $\phantom{\mbox{if}}$ $(\au, \sigma, \sigma''') \in A_2$ and $(\at, \sigma''', \sigma'') \in A_1$.\\
    \end{tabular}  &
    \includegraphics[width=0.45\textwidth,valign=M]{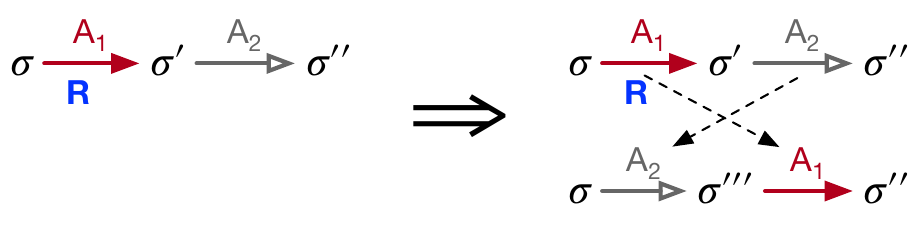} \\

    \begin{tabular}{ll}
      ~\\
    (2)
    & if $M(A_1,\at, \sigma) \sqsubseteq \MN$ and $(\at, \sigma, \sigma') \in A_1$ and \\
    & $\phantom{\mbox{if}}$ $M(A_2, \au, \sigma') \sqsubseteq \ML$ and $(\au, \sigma', \sigma'') \in A_2$, \\
    & then there exists $\sigma'''$ such that \\
      & $\phantom{\mbox{if}}$ $(\au, \sigma, \sigma''') \in A_2$ and $(\at, \sigma''', \sigma'') \in A_1$.\\
    \end{tabular}  &
    \includegraphics[width=0.45\textwidth,valign=M]{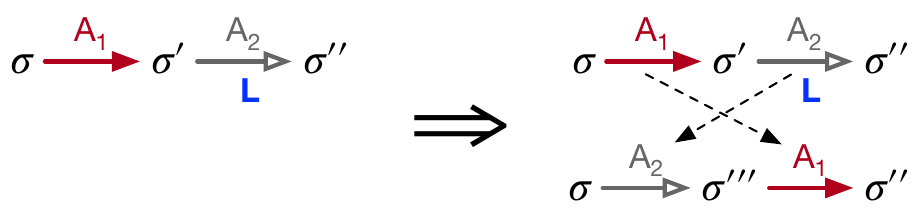}     \\
  \end{tabular}
  \end{center}
  \begin{center}
  \begin{tabular}{l@{\qquad}c}
    \begin{tabular}{ll}
      ~\\
    (3)
      & if $M(A_1,\at, \sigma) \sqsubseteq \MN$ and $(\at, \sigma, \sigma') \in A_1$ and\\
      & $\phantom{\mbox{if}}$  $M(A_2, \au, \sigma) = e$ for some $e$, \\
      & then $M(A_2, \au, \sigma') = e$. \\
    \end{tabular}  &
    \includegraphics[width=0.45\textwidth,valign=M]{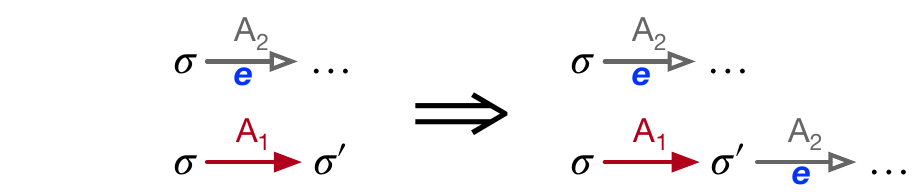}  \\
    \begin{tabular}{ll}
      ~\\
    (4)
      & if $M(A_1,\at, \sigma) \sqsubseteq \MN$ and $(\at, \sigma, \sigma') \in A_1$ and\\
      & $\phantom{\mbox{if}}$  $M(A_2, \au, \sigma) \sqsubseteq \ML$ and $(\au, \sigma, \sigma'') \in A_2$, \\
      & then there exists $\sigma'''$ such that \\
      & $(\au, \sigma', \sigma''') \in A_2$ and $(\at, \sigma'', \sigma''') \in A_1$.
    \end{tabular}  &
    \includegraphics[width=0.45\textwidth,valign=M]{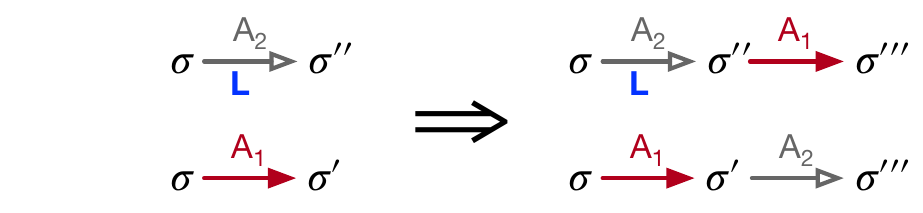} 
  \end{tabular}
  \end{center}  
\end{definition}


\myclearpage
\subsection{Correctness}

The central correctness theorem for {\logic} is that verified
programs do not go wrong by, for example, failing an assertion. 

\begin{theorem}[Soundness]
  \label{thm:sound}
  If $\vdash \Sigma$ then $\Sigma$ does not go wrong.
\end{theorem}
The proof 
appears in full in the \ecoopOrExtended{extended version of this paper~\cite{XXX}}{Appendix}. The basic  structure is as follows.
\begin{enumerate}
  
\item
We first develop an instrumented semantics that enforces
the mover specification $M$ and also that each thread consists
of reducible sequences separated by yields.

\item
In addition to the usual preemptive scheduler, we also develop a
non-preemptive scheduler for the instrumented semantics that context
switches only at $\tyield$s.

\item
We show that the instrumented semantics under the preemptive scheduler
behaves the same as the standard semantics except that it may go wrong
more often.

\item
We use a reduction theorem to show that programs exhibit
the same behavior under the preemptive and non-preemptive instrumented semantics.

\item
Finally, we use a preservation
argument~\cite{DBLP:journals/iandc/WrightF94} to show that verified
programs do not go wrong under the non-preemptive instrumented
semantics.

\item The steps above then imply that verified programs do not go
  wrong under the preemptive standard semantics.
\end{enumerate}

\myclearpage
\section{Related Work}
\label{sec:related}

\subsection*{Modular Reasoning}

Concurrent software verification introduces a number of scalability
challenges that require a synthesis of various notions of modularity
or abstraction to address.
For example, \emph{procedure-modular} reasoning tackles large code
bases by verifying each procedure with respect to a specification of
other procedures in the system.
Rely-guarantee logic~\cite{DBLP:journals/toplas/Jones83} augments
procedure-modular reasoning with a notion of \emph{thread-modular} reasoning that
accommodates multiple threads by verifying each thread
with respect to a specification of other threads in the system.
As demonstrated in Section~\ref{sec:overview}, systems like RG logic
that support procedure-modular and thread-modular reasoning have great
potential, but they are limited by entanglement between library 
and client specifications.

To address that limitation, mover logic augments procedure-modular and
thread-modular reasoning with Lipton's theory of
reduction~\cite{DBLP:journals/cacm/Lipton75}.  This complementary form
of ``interleaving'' modularity limits the number of interleavings
that must be considered and enables more precise procedure
specifications for atomic functions.

In other work, separation logic combines procedure-modular reasoning
with a notion of \emph{heap-modular}
reasoning~\cite{DBLP:conf/csl/OHearnRY01,DBLP:conf/lics/Reynolds02},
which enables verification of sub-goals while ignoring irrelevant heap
objects.  Separation logic has been the foundation for a variety of
verification
tools~\cite{DBLP:conf/fmco/BerdineCO05,DBLP:conf/nfm/JacobsSPVPP11,DBLP:series/natosec/0001SS17}.
Concurrent separation logics including, for
example~\cite{DBLP:conf/concur/VafeiadisP07,DBLP:journals/tcs/OHearn07,DBLP:journals/tcs/Brookes07,DBLP:phd/ethos/Vafeiadis08},
extend those ideas to a concurrent setting.  While initially focused
on noninterference via disjoint access and read-only sharing, later
work~\cite{DBLP:conf/esop/DoddsFPV09,DBLP:conf/popl/Dinsdale-YoungBGPY13}
supports more tightly-coupled threads.

Much of the work on concurrent separation logic focuses on
\emph{resources} (\emph{e.g.}, heap locations) and on ensuring threads access
disjoint resources (hence ensuring noninterference). In contrast,
mover logic focuses on commuting \emph{actions}.

Concurrent separation logic and mover logic also differ in where
thread interference specifications are placed.  Concurrent separation
logic conveniently merges interference (or resource footprint)
specifications into each method's precondition, thus enabling the
logic to capture sophisticated resource usage idioms in a concise and
elegant manner.  Deny-guarantee
reasoning~\cite{DBLP:conf/esop/DoddsFPV09} extends concurrent
separation logic to focus more on actions rather than resources.  In
particular, a method's precondition can include an ``action map''
specifying what actions the method (and its concurrent threads) may
perform.  This action map is analogous to our mover specifications.
Several projects employ permissions or ownership, similar to
separation logic, to reason about which memory locations are available
to different threads. These include
Viper~\cite{DBLP:conf/vmcai/0001SS16} and
VerCors~\cite{DBLP:conf/ifm/BlomDHO17}.  These systems do not support reduction.

An important topic of for future study is how to extend mover logic
with a notion of heap modularity, perhaps similar to the core ideas of
concurrent separation logic or dynamic
frames~\cite{DBLP:conf/ecoop/BanerjeeNR08,DBLP:conf/fase/SmansJPS08,DBLP:conf/fm/Kassios06}.
This body of work may also provide insight into how to develop a
compositional semantics based on mover logic.

\subsection*{Reduction-based Techniques}

QED~\cite{DBLP:conf/icse/Elmas10} is a program calculus and
verification procedure for concurrent programs.  It utilizes iterative
reduction and abstraction refinement to increase the size of the
blocks that can be considered serializable regions (at the abstract
level).  That approach has been shown to be quite successful for
verifying complex concurrent code and has inspired a number of
subsequent verification tools described below.
Mover logic is a 
complementary approach in that the
combination of RG reasoning and reduction enables direct verification of code
with \tyield{} points, without the need to create
layers of abstractions.  As part of that, mover logic supports
specifying and reasoning about functions that are not atomic, which is
not supported in QED.  We also note that QED checks the commutativity
properties of an action via a pairwise check with all other actions in
the code, whereas mover logic uses the mover specification
validity check for that purpose.

Several more recent verification tools utilize the same approach of
writing a series of programs related by refinement, abstraction, and
reduction.  These include the CIVL verifier~\cite{DBLP:conf/cav/HawblitzelPQT15,DBLP:conf/fmcad/KraglQ21,DBLP:conf/cav/KraglQH20,DBLP:conf/pldi/KraglEHMQ20,DBLP:conf/cav/KraglQ18,DBLP:conf/concur/KraglQH18}
and the Armada verifier~\cite{DBLP:conf/pldi/LorchCKPQSWZ20}.
They are capable of handling sophisticated concurrent code, but do
require the programmer to write and maintain multiple versions of the
source code.
The correctness arguments for these tools have typically been based
on monolithic proofs.  

%
Calvin-R~\cite{DBLP:journals/jot/FreundQ04} developed a number of
early ideas related to reduction and
thread-modular reasoning.  The \textsc{Anchor}
verifier~\cite{DBLP:journals/pacmpl/FlanaganF20} builds on ideas behind
Calvin-R and CIVL to create a verification technique supporting an
executable, object-oriented target language, a variety of
synchronization primitives, and a new notation for specifying the
interference between threads that is the foundation for our mover
specifications.  While effective at some verification tasks, {\sc
  Anchor}'s correctness arguments are also challenging to understand
and build upon.
Further, {\sc Anchor} is inherently limited to small
programs because it inlines nested calls during verification, with no
mechanism for procedure-modular reasoning.  Mover logic
may provide a useful foundation for a procedure-modular extension of
{\sc Anchor}.

The difficulty in assessing the strengths and weaknesses of the tools
mentioned above without a robust underlying logic capturing what they
do inspired this work.  Mover logic may provide such a foundation,
detached from any particular full-scale implementation, that it is
accessible, general, and extensible.  We hope implementations based
on mover logic will follow, as the logic clarifies exactly what
conditions must be met in reduction-based verifiers that attempt to
integrate modular reasoning in the presence of interference.


\subsection*{Coq-based Techniques}
Complementary approaches develop proof frameworks for verifying
concurrent programs in Coq~\cite{coq}.
For example, CCAL~\cite{DBLP:conf/pldi/GuSKWKS0CR18} provides a
compositional semantic model for composing and verifying the
correctness of multithreaded components. CCAL focuses on only
rely-guarantee reasoning~\cite{DBLP:journals/toplas/Jones83} and not
reduction.
CSpec~\cite{DBLP:conf/osdi/ChajedKLZ18} is a Coq library for verifying
concurrent systems modeled in Coq~\cite{coq} using movers and
reduction. While highly expressive, particularly because additional
proof techniques can be added as additional Coq code, users must write
significant Coq code for both specifications and proofs to use such a
system.  We have focused on a logic more amenable to fully automatic
reasoning.
Iris~\cite{DBLP:journals/jfp/JungKJBBD18} uses higher-order separation
logic to verify correctness of higher-order imperative programs.

\subsection*{Model Checking}

An orthogonal approach to software verification utilizes explicit
state, exhaustive model checking.  Such approaches have lower
programmer overhead than other techniques, but they are
non-modular~\cite{DBLP:conf/icalp/EmersonC80,DBLP:conf/lop/ClarkeE81,DBLP:journals/toplas/ClarkeES86}.
Specialized techniques, including
reduction~\cite{DBLP:conf/vmcai/HatcliffRD04} and partial-order
methods~\cite{DBLP:conf/lics/GodefroidW91,DBLP:conf/popl/Godefroid97,DBLP:conf/cav/Peled94},
have been used to limit state-space explosion while checking
concurrent programs.  A variety of concurrent software model
checkers~\cite{DBLP:conf/issta/ChamillardC96,DBLP:conf/popl/Yahav01,DBLP:conf/osdi/MusuvathiQBBNN08}
have demonstrated the potential of these approaches in
constrained settings.

\myclearpage
\section{Summary}
\label{sec:summary}

Over the last two decades, 
several promising multithreaded program
verifiers have leveraged reduction to 
verify sophisticated concurrent code including
non-blocking algorithms, dynamic data race detectors, and garbage
collectors by leveraging  precise,
reusable specifications for atomic functions.
The reasoning used by these verifiers, 
including the notion of which programs are verifiable,
and why the verification process is sound,
is unfortunately rather complex.
In contrast, Hoare logic~\cite{DBLP:journals/cacm/Hoare69} provides an accessible foundation for sequential verifiers, and RG
logic~\cite{DBLP:journals/toplas/Jones83} provides a similar
foundation for some multithreaded verifiers.

In developing mover logic, we aim to facilitate future research on 
reduction-based verification. 
Mover logic provides a declarative and formal explanation of reduction-based verification, making it easier to understand 
which programs are verifiable, or not, and why; 
which functions can be specified as atomic; 
what atomic and non-atomic function specifications mean; 
which code blocks are reducible; 
where yield annotations are required, etc.
The correctness proof for a reduction-based verifier need only show that the verifier follows the
rules of mover logic, a significant simplification over existing proof
techniques. 

We hope that
mover logic inspires the development of more expressive reduction-based  logics and
 verification tools, potentially supporting  features such as objects, data abstraction, dynamic allocation,
dynamic thread creation, and precise frame
conditions~\cite{DBLP:conf/ecoop/BanerjeeNR08,DBLP:conf/fase/SmansJPS08,DBLP:conf/fm/Kassios06}.


\bibliography{pruned.bib}

\ecoopOrExtended{\end{document}